\documentclass[12pt]{iopart}
\usepackage[utf8]{inputenc}
\usepackage{graphicx}
\usepackage{color}
\usepackage{hyperref}
\usepackage{bm}
\usepackage{amssymb}
\usepackage{amsmath}
\usepackage{ulem}

\newcommand{\diff}{\mathrm{d}}
\newcommand{\p}{\partial}
\newcommand{\img}{\mathrm{i}}

\newcommand{\ele}{\mathrm{e}}
\newcommand{\ion}{\mathrm{i}}
\newcommand{\vthe}{v_{\mathrm{th,e}}}

\newcommand{\by}{B_y^{\mathrm{eq}}}

\newcommand{\apt}{\tilde{A}_\para}

\newcommand{\ept}{\tilde{E}_\para}
\newcommand{\vx}{\tilde{v}_x}

\newcommand{\omegar}{\Re(\omega)}
\newcommand{\omegai}{\Im(\omega)}
\newcommand{\omegat}{\omega_{\ast T_{0\mathrm{e}}}}
\newcommand{\omegan}{\omega_{\ast n_{0\mathrm{e}}}}
\newcommand{\tauA}{\tau_{\mathrm{A}}}
\newcommand{\tauR}{\tau_{\mathrm{R}}}
\newcommand{\teq}{\mathrm{eq}}
\newcommand{\para}{\parallel}

\newcommand{\agk}{\texttt{AstroGK}}
\newcommand{\at}{\alpha_{\mathrm{t}}}
\newcommand{\att}{\alpha_{\mathrm{tt}}}
\newcommand{\attt}{\alpha_{\mathrm{t}}\alpha_{\mathrm{tt}}}
\newcommand{\ai}{\alpha_{\mathrm{i}}}
\newcommand{\atti}{\alpha_{\mathrm{tt,i}}}

\newcommand{\RN}[1]{{\color{black}{#1}}}
\newcommand{\MY}[1]{{\color{black}{#1}}}

\begin{document}
\title[Destabilization mechanism of the microtearing mode]
{Destabilization mechanism of the collisional microtearing mode in magnetized slab plasmas }

\author{Mitsuyoshi Yagyu$^1$ \& Ryusuke Numata$^2$}
\address{$^1$Graduate School of Simulation Studies, University of Hyogo, 
7-1-28 Minatojima Minami-machi, Chuo-ku, Kobe, Hyogo 650-0047, Japan}
\address{$^2$Graduate School of Information Science, University of Hyogo, 
7-1-28 Minatojima Minami-machi, Chuo-ku, Kobe, Hyogo 650-0047, Japan}
\ead{mitsu.yagyu@gmail.com}
\vspace{10pt}

\begin{abstract}
The destabilization mechanism of the collisional microtearing mode driven by an electron temperature gradient is studied using theoretical analyses and gyrokinetic simulations including a comprehensive collision model, \RN{in magnetized slab plasmas}.
The essential destabilization mechanism of the microtearing mode is the lag of the parallel inductive electric field behind the magnetic field owing to the time-dependent thermal force and inertia force induced by the velocity-dependent electron--ion collisions. Quantitative measurements of the collision effects enable us to identify the unstable regime against collisionality and reveal the relevance of the collisional microtearing mode with existing toroidal experiments. A nonlinear simulation demonstrates that the microtearing mode does not drive magnetic reconnection with the explosive release and conversion of the magnetic energy.
\end{abstract}
%
\vspace{2pc}
\noindent{\it Keywords}: microtearing mode, time-dependent thermal force, electromagnetic drift wave, self-filamentation mode
%
%
%
%

\section{Introduction}

Micro-instabilities driven by electron temperature gradients excite turbulence and lead to anomalous electron thermal transport. Because such a transport event is associated with the degradation of confinement in fusion devices, predicting the transport level is one of the most important issues in designing future fusion devices. However, the physics of electron thermal transport in fusion devices is not yet fully understood. Electrostatic turbulence driven by the electron temperature gradient (ETG) mode or trapped electron mode (TEM) is known to increase transport in low-$\beta$ devices ($\beta$ is the ratio of the plasma pressure to the magnetic pressure). In higher-$\beta$ devices, turbulence driven by electromagnetic micro-instabilities involving magnetic fluctuations may play a role in causing anomalous transport.

A number of gyrokinetic simulations dedicated to specific experimental devices, such as  standard tokamaks~\cite{doerk2012, hatch2016}, spherical tokamaks~\cite{applegate2007, wong2007,guttenfelder2011, moradi2013}, and reversed-field pinches (RFP)~\cite{predebon2013, carmody2013}, have demonstrated that the microtearing mode, which is an electromagnetic mode, becomes unstable and causes electron transport comparable to the measured experimental value~\cite{guttenfelder2012PoP, jian2019}. The microtearing mode enables topological changes in magnetic field structures and generates small-scale islands that may overlap to provide a stochastic magnetic field structure, thereby enhancing electron transport along the field lines~\cite{doerk2011, rechesterrosenbluth1978}.
Nonlinear gyrokinetic simulations for the National Spherical Torus Experiment (NSTX) device~\cite{guttenfelder2012PoP, kaye2007} have estimated that although an oversimplified assumption that local transport and global confinement can be directly compared is employed, transport scaling with respect to collisionality is roughly consistent with global energy confinement scaling.

The first linear theory of microtearing instability was presented by Hazeltine et al.~\cite{hazeltine1975} using a kinetic model that included the rigorous Fokker--Planck collision operator. It extends the tearing mode theory based on a fluid model~\cite{fkr1963} to include a non-uniform background temperature and to avoid assuming a highly collisional fluid limit. In these analyses, a standard multi-scale treatment was employed. Given a sheared magnetic field $\bm{B}(x)$, a boundary layer (inner region) may appear near the singular point $x=x_{\mathrm{s}}$, where $\left.\bm{B} \cdot \nabla\right|_{x_{\mathrm{s}}}=0$ owing to the singular perturbation nature of non-ideal terms, such as resistivity. The inner-region solution is asymptotically matched with the outer-region solution of an ideal fluid model. Hazeltine et al. discovered that the velocity dependence of the collision frequency is indispensable for tapping the free energy associated with the temperature gradient to drive microtearing instability.
Subsequently, the electron--ion collision frequency dependence of the microtearing mode has been investigated theoretically~\cite{drake1977} and numerically~\cite{dippolito1980, gladd1980}. A remarkable conclusion has demonstrated that the growth rate of the mode is non-monotonic with respect to the collision frequency, with the growth rate peaking at $\omega_{\ast\ele} \lesssim \nu_{\ele\ion}$, where $\omega_{\ast\ele}$ and $\nu_{\ele\ion}$ are the electron diamagnetic drift frequency and electron--ion collision frequency, respectively. The microtearing mode is stabilized in both collisionless and collisional limits.
The collisional driving mechanism of the microtearing mode is ascribed to the thermal force arising from the interaction between the electron temperature gradient and velocity-dependent collisions. Hassam derived electron transport equations using the Chapman--Enskog expansion of the electron Fokker--Planck equation under the assumption that $\omega/\nu_{\ele\ion}\ll 1$~\cite{hassam1980_1,hassam1980_2}, where the second-order terms in $\omega/\nu_{\ele\ion}$ are retained. ($\omega^{-1}$ is the timescale of interest.) It is shown that the time-dependent thermal force appears only in the second order and makes the mode unstable.

However, the linear theory of the microtearing mode remains controversial. The most debatable issue is whether the microtearing mode becomes unstable in weakly collisional or collisionless plasmas.
Some numerical evidence shows a reconnecting instability driven by the electron temperature gradient in collisionless slab plasmas~\cite{kobayashi2014,zocco2015,geng2020}. It is argued that there is a tearing-parity electromagnetic ETG mode in a slab, which has different characteristics from \RN{the} microtearing \RN{mode}~\cite{zocco2015,geng2020}.
In toroidal geometry, theoretical and simulation studies~\cite{catto1981,dickinson2013,chandran2022} have shown that trapped particle effects can drive the collisionless microtearing mode. 

In this paper, we review existing theoretical studies on the microtearing mode and present numerical analyses using a gyrokinetic model, focusing on collision effects. We restrict our analyses to slab geometry as the simplest problem setup to clarify the essential destabilizing mechanism of the mode. 
Through the analyses, we conclude that the slab microtearing mode is stable in the collisionless regime \MY{as long as a certain condition for $\beta$ and $\nu_{\ele\ion}$ is satisfied.}

The remainder of this paper is organized as follows:
We present theoretical analyses of tearing-type instability using a fluid model in Sec.~\ref{sec:thore_analysis}. The effects of the magnetic shear and temperature gradient are highlighted.
Numerical simulations of the microtearing mode using the electromagnetic gyrokinetic simulation code $\agk$~\cite{numata2010} are presented in Sec.~\ref{sec:simulation}. 
The collision effects are estimated in terms of the transport coefficients, and an instability region is identified where the time-dependent thermal force works as an instability mechanism. A nonlinear evolution of the microtearing mode is also discussed in the same section. 
Finally, we summarize the paper in Sec.~\ref{sec:summary}.

\section{Linear theoretical analysis using a fluid model}
\label{sec:thore_analysis}
Tearing-type instability analyses are formulated using Ohm's law and the ion vorticity equation. They determine the evolution of the electromagnetic fluctuations $A_{\parallel}$ and $\phi$ and are coupled through the conductivity $\sigma_{\parallel}=j_{\parallel}/E_{\parallel}$, where $A_{\parallel}$, $j_{\parallel}$, and $E_{\parallel}$ are the magnetic vector potential, current density, and electric field along the magnetic field, respectively, and $\phi$ denotes the electrostatic potential. The form of conductivity varies depending on the model. In the following analyses, we use collisional conductivity derived from the fluid model of Hassam~\cite{hassam1980_1}. This is essentially the same as that derived from the kinetic models, as long as the collision effects are sufficiently strong.

In this section, we review instability analyses induced by the temperature gradient and collisions. We first consider the case without magnetic shear, where the electromagnetic drift-wave-type mode is destabilized owing to time-dependent thermal force.
Then, we proceed to tearing-type instabilities in non-uniform temperature plasmas. The similarity between the instability mechanisms with and without magnetic shear is discussed.


\subsection{Thermal force in a time-dependent magnetic field}
\label{subsec:self-filamentation}
The thermal force is an effective force acting on electrons with a non-uniform temperature profile. Because the collisionality $\nu_{\ele\ion}$, defined as (see, for example,~\cite{helandersigmar})
\begin{equation}
    \nu_{\ele\ion} \equiv \frac{\sqrt{2} n_{\ion} Z^{2} e^{4} \ln \Lambda}{16 \pi \epsilon_{0}^{2} m_{\ele}^{1/2} T_{\ele}^{3/2}},
\label{eq: collision_freq_definition}
\end{equation}
depends on the temperature, an imbalance of friction forces caused by collisions with background ions occurs, which leads to net momentum transfer in the opposite direction to the temperature gradient. In the presence of a strong magnetic field ($\nu_{\ele\ion}/\Omega_{\mathrm{ce}} \ll 1$, where $\Omega_{\mathrm{ce}}$ is the electron cyclotron frequency), electrons move freely along the field lines and gyrate around the magnetic field. Consequently, the thermal force in the directions parallel and perpendicular to the magnetic field differs by a factor of $\nu_{\ele\ion}/\Omega_{\mathrm{ce}}$. In the following, we only consider the thermal force in the parallel direction. The thermal force for strongly magnetized plasmas is given by
\begin{align}
    F^{\mathrm{th}}_{\para} = - \at n_{\ele} \nabla_{\para} T_{\ele},
\end{align}
where $n_{\ele}$ and $T_{\ele}$ are the electron density and temperature, respectively, and the numerical factor $\at$ was estimated as $\at=0.71$ for hydrogen by Braginskii~\cite{braginskii1965}. The derivative is taken in the direction of the magnetic field $\nabla_{\para} = \bm{B}\cdot\nabla/|\bm{B}|$.

Hassam~\cite{hassam1980_1} later obtained nonlinear transport equations for electrons with a less restricted constraint than that of Braginskii's, that is, $\omega \ll \nu_{\ele\ion}$.
The generalized Ohm's law is given by\RN{~\cite{hassam1980_2}}
\begin{equation}
   \ai \frac{\p}{\p t} (m_{\ele} n_{\ele} u_{\para,\ele}) 
        =
    - e n_{\ele} E_{\para} - \nabla_{\para} (n_{\ele} T_{\ele}) + \e n_{\ele} \eta j_{\para}
    - \at n_{\ele} \nabla_{\para} T_{\ele}
    + \at \att \frac{1}{\nu_{\ele\ion}} \frac{\p}{\p t}
   \left(n_{\ele} \nabla_{\para} T_{\ele} \right).
   \label{eq:generalized_ohm}
\end{equation}
The resistivity $\eta$ is also written in terms of the collisionality $\eta/\mu_{0} = \alpha_{\eta} \nu_{\ele\ion} d_{\ele}^{2}$, where $\mu_{0}$ is the vacuum permeability, and $d_{\ele}$ is the electron skin depth. The coefficient is $\alpha_{\eta}=0.38$~\cite{braginskii1965, spitzer1953}\footnote{Note that the number is different from that written in~\cite{braginskii1965} because we use $\nu_{\ele\ion}=(3\sqrt{\pi}/4) /\tau_{\ele}$, where $\tau_{\ele}$ is the collision time used in~\cite{braginskii1965}.}.
The newly appeared smaller-order terms in $\omega/\nu_{\ele\ion}$ are the inertia term on the LHS ($m_{\ele}$ is the electron mass, and $u_{\para,\ele}$ is the electron flow in the magnetic field direction) and the time-dependent thermal force in the last term of the RHS. $\at,\att$, and $\ai$ are numerical factors depending on the collision model, which vanish in the collisionless limit. It will be shown shortly that the time-dependent thermal force brings about an instability where a magnetic field fluctuation is amplified to develop current filaments (the {\it self-filamentation} mode)~\cite{hassam1980_2}.

\begin{figure}[ht]
\begin{center}
	\includegraphics[width=100mm]{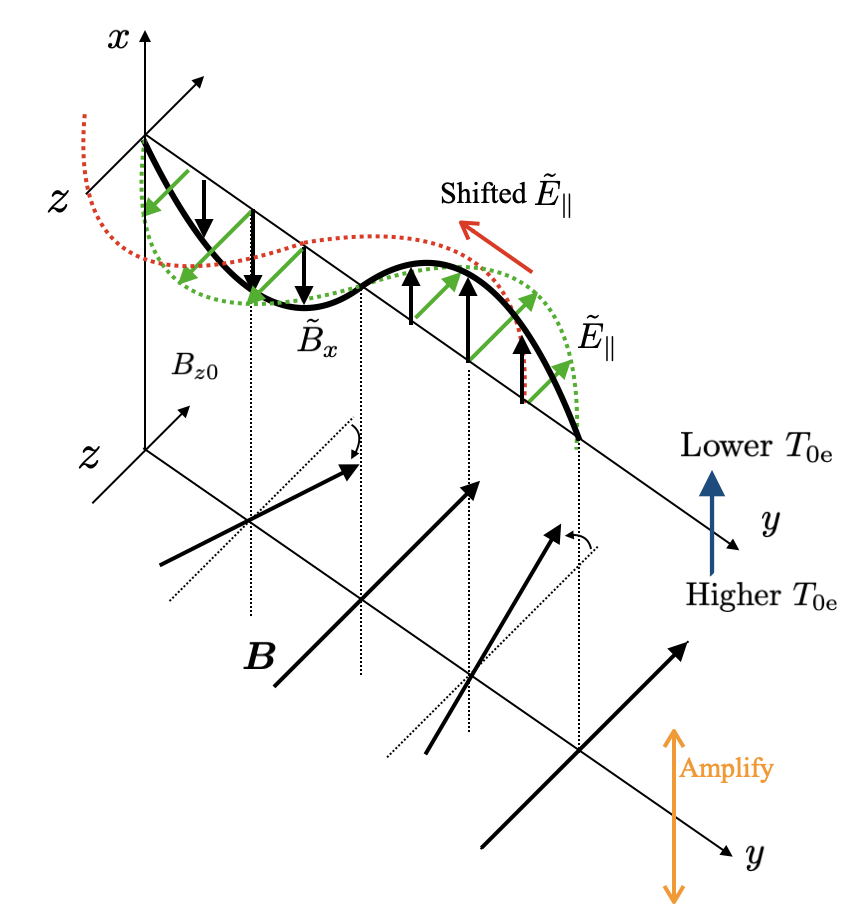}
\end{center}
	\caption{Schematic illustration of the electromagnetic drift wave. The given perturbation $\tilde{B}_x$ corresponding to the black solid line generates the parallel electric field represented by the green dashed line. In the bottom, the magnetic field $\bm{B} =\tilde{B}_x\hat{\bm{x}}+B_{z0}\hat{\bm{z}}$ oscillation is shown. If there is no phase shift of the parallel electric field $\ept$, the amplitude of the magnetic field does not grow. The lag of $\ept$ owing to the time-dependent thermal force denoted by the red arrow is necessary for the self-filamentation instability.}
\label{fig:emdrift_filamentation}
\clearpage
\end{figure}

We consider a non-uniform plasma with density $n_{0\ele}(x)$ and temperature $T_{0\ele}(x)$ embedded in a uniform magnetic field given by $\bm{B}_{0}=B_{z0} \hat{\bm{z}}$. The density and temperature gradients are taken in the $x$ direction: $L_{n_{0\mathrm{e}}}^{-1} = - \p_{x} \ln n_{0\mathrm{e}}$ and $L_{T_{0\mathrm{e}}}^{-1} = - \p_{x} \ln T_{0\mathrm{e}}$. When a small perturbation $\tilde{\bm{B}} = \tilde{B}_{x} \hat{\bm{x}} = \partial_{y} \tilde{A}_{z}\hat{\bm{x}}$ ($\tilde{B}_{x} \ll B_{z0}$) is applied, it propagates because of the electron force balance along the field line. Faraday's law in the parallel direction,
\begin{align}
    \frac{\p \apt}{\p t} & = - \ept,
\end{align}
and the generalized Ohm's law \eqref{eq:generalized_ohm} constitute a closed set of equations for magnetic fluctuation. Note that $\apt = \tilde{A}_{z}$ if $B_{z0}=\p_{x}A_{y0}$. By eliminating the electric field and assuming a plane-wave solution $\propto \mathrm{e}^{-\img(\omega t - k_{y} y)}$, we obtain
\begin{align}
    - \img \omega \apt = - \img \ai \omega \frac{m_{\ele}\tilde{u}_{\para,\ele}}{e}
    - \eta \tilde{j}_{\para} - \frac{\tilde{B}_{x}}{B_{z0}} \frac{T_{0\ele}}{e}
        \left[L_{n_{0\ele}}^{-1} + L_{T_{0\ele}}^{-1}
        \left(1 + \at  + \at \att \frac{\img \omega}{\nu_{\ele}} \right)
        \right].
\end{align}
The current and parallel electron flow are expressed in terms of $\tilde{A}_{\para}$ if the ion parallel flow is negligible, $\tilde{j}_{\para} = - e n_{0\ele} \tilde{u}_{\para,\ele} = - (1/\mu_{0}) \nabla^{2} \apt$. By introducing the drift frequency,
\begin{align}
    \omega_{\ast n_{0\ele}} & = \frac{k_{y}\rho_{\ele}}{2} \frac{\vthe}{L_{n_{0\ele}}}, \\
    \omega_{\ast T_{0\ele}} & = \frac{k_{y}\rho_{\ele}}{2} \frac{\vthe}{L_{T_{0\ele}}},
\label{eq:diamag_frequency}
\end{align}
where the thermal speed and Larmor radius are respectively defined by $\vthe = \sqrt{2T_{0\ele}/m_{\ele}}$ and $\rho_{\ele}=\sqrt{2m_{\ele}T_{0\ele}}/(eB_{z0}) = \vthe/\Omega_{\mathrm{ce}}$, the dispersion relation becomes
\begin{align}
    \left(
        1-\img \attt \frac{\omegat}{\nu_{\ele\ion}}
    \right)\frac{\omega}{\nu_{\ele\ion}}
    =
    \frac{\omegan+(1+\at)\omegat}{\nu_{\ele\ion}} - \img \alpha_{\eta}(k_y d_{\ele})^2
    \left(
        1-\frac{\ai}{\alpha_{\eta}}\frac{\img \omega}{\nu_{\ele\ion}}
    \right).
\label{eq:sfm_disp}
\end{align}
We seek a solution that satisfies the condition,
\begin{align}
    \frac{\omega}{\nu_{\ele\ion}}
    \sim \frac{\omega_{\ast n_{0\ele}}}{\nu_{\ele\ion}}
    \sim \frac{\omega_{\ast T_{0\ele}}}{\nu_{\ele\ion}}
    \sim (k_{y} d_{\ele})^{2}
    \sim \epsilon \ll 1,
\end{align}
and find the solution
\begin{align}
    \Re(\omega) &= (\omegan+(1+\at)\omegat) (1- \ai (k_y d_{\ele})^2) + \alpha_{\eta} \nu_{\ele\ion}(k_y d_{\ele})^2 \attt\frac{\omegat}{\nu_{\ele\ion}}, 
\label{eq:sfm_real_freq}
    \\
    \Im(\omega) &= (\omegan+(1+\at)\omegat) \attt\frac{\omegat}{\nu_{\ele\ion}}-\alpha_{\eta}\nu_{\ele\ion}(k_y d_{\ele})^2 (1 - \ai (k_y d_{\ele})^2).
\label{eq:sfm_growth_rate}
\end{align}
In the leading order of $\epsilon$, the fluctuation propagates in the $y$ direction with a frequency of $\omega = \omega_{\ast n_{0\ele}} + (1+\alpha_{\mathrm{t}}) \omega_{\ast T_{0\ele}}$. We refer to this wave as an electromagnetic drift wave. In the subsequent order, a positive imaginary part of $\omega$ arises. Only the time-dependent thermal force proportional to $\att$ has a destabilizing effect. The electromagnetic drift wave and its destabilization are shown in Fig.~\ref{fig:emdrift_filamentation}. The time-dependent thermal force causes a lag in the inductive electric field behind the magnetic field perturbation and leads to an increase in the perturbation.

\subsection{Analysis of collisional microtearing instability}
\label{subsec:tearing_analysis}
\begin{figure}[ht]
\begin{center}
	\includegraphics[width=90mm]{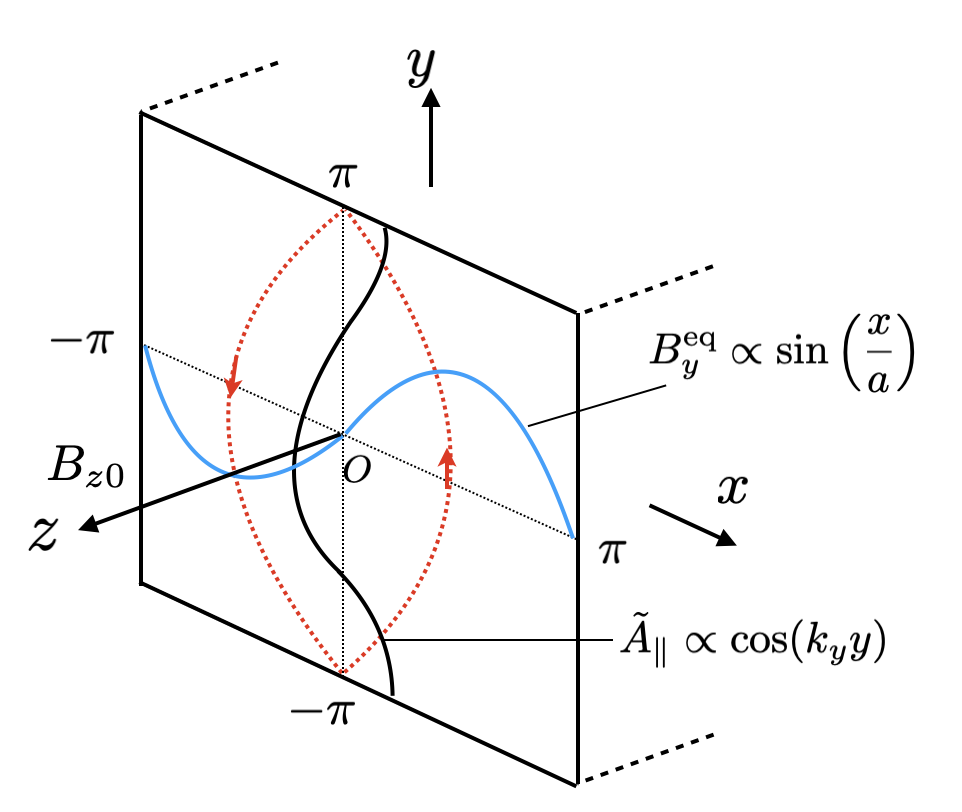}
\end{center}
	\caption{Illustration of the initial magnetic fields in the two-dimensional slab.
	The black and blue solid lines are the perturbed and equilibrium magnetic fields, respectively.
	The red dashed line is the resultant magnetic configuration with the X-point formed at $(x,y)=(0,\pm \pi)$.}
\label{fig:slab}
\clearpage
\end{figure}

In addition to the configuration discussed in the previous section, we apply a shear to the equilibrium magnetic field. The total equilibrium magnetic field is given by $\bm{B}_{0} = B_{z0} \hat{\bm{z}} + B_{y}^{\mathrm{eq}}(x) \hat{\bm{y}}$ ($B_{z0} \gg \by(x))$, where $B_{y}^{\teq}(x)=-\p
A_{\para}^{\teq}(x)/\p x$ is the reconnecting component of the magnetic field. We choose
\begin{equation}
    \by(x) = B_{y0} \sin \left( \frac{x}{a} \right),
\label{eq:by}
\end{equation}
where $B_{y0}$ and $a$ denote the amplitude and scale length of $B_{y}^{\mathrm{eq}}$, respectively. We apply a small perturbation $\tilde{\bm{B}}=\tilde{B}_x \hat{\bm{x}}=\p_y \apt \hat{\bm{x}}$ ($\by(x) \gg \tilde{B}_x$). Figure~\ref{fig:slab} illustrates the initial magnetic field geometry in the slab. Because of the additional magnetic field component, the evolution of $\apt$ now couples with plasma flow, as described by Faraday's law,
\begin{equation}
    \frac{\p \apt}{\p t} 
    = - \ept - \nabla_{\para} \tilde{\phi}
    = -\ept - \img k_{\para} \tilde{\phi},
\label{eq:faraday_tearing}
\end{equation}
where $k_{\para} = k_y B_{y}^{\mathrm{eq}}/B_{z0}$. Note that the flow in the plane perpendicular to the magnetic field $\bm{B}_{0}$ is given by the $\bm{E}\times\bm{B}$ flow, namely, $\tilde{\bm{v}} = -\nabla \tilde{\phi} \times \hat{\bm{z}}/B_{z0}$ ($\tilde{\phi}$ is the electrostatic potential). The vorticity equation for ions\footnote{The ion diamagnetic drift may be included.} demands that
\begin{equation}
    \frac{\p}{\p t}\left(\frac{\p^2}{\p x^2}-k_y^2 \right) \frac{\tilde{\phi}}{B_{z0}}
    =
    - \img k_y V_{\mathrm{A}}^2\frac{{\by}}{B_{y0}^2}
    \left[\left(\frac{\p^2}{\p x^2}-k_y^2 \right)\apt - \frac{{\by}''}{\by}\apt \right],
\label{eq:vorticity_tearing}
\end{equation}
where $V_{\mathrm{A}} = B_{y0}/\sqrt{\mu_{0}m_{\ion} n_{0\mathrm{i}}}$ is the Alfv\'en velocity. Ohm's law determining $\ept$ also includes $k_{\para}$ and therefore couples with density and temperature perturbations. These effects are encapsulated in the conductivity $\sigma_{\para} = \tilde{j}_{\parallel}/\ept$:
\begin{multline}
    \sigma_{\parallel} 
    = \frac{1}{\eta}
    \left[
        1-\frac{\omegan}{\omega}-\frac{\omegat}{\omega}\frac{1+\at+\img \attt \omega/\nu_{\ele\ion}}{1+\img k_{\para}^2 \kappa_{\para}/\omega}
    \right] \\
    \times
    \left[
        1-\frac{\ai}{\alpha_{\eta}}\frac{\img \omega}{\nu_{\ele\ion}}+
        \img\frac{1}{\alpha_{\eta}}\frac{k_{\para}^2 D_{\para}}{\omega}
        \left(
            1+\frac{2}{3}\frac{(1+\at) (1+\at + \img \attt\omega/\nu_{\ele\ion})}{1+\img k_{\para}^2 \kappa_{\para}/\omega}
        \right)
    \right]^{-1},
\label{eq:generalized_sigma}
\end{multline}
where $D_{\parallel}\equiv \vthe^{2}/(2\nu_{\ele\ion})$ and $\kappa_{\para} \equiv (2/3) \alpha_{\kappa} D_{\para}$ are the parallel diffusion coefficient and parallel thermal conductivity, respectively. The coefficient is $\alpha_{\kappa}=4.2$~\cite{braginskii1965}. We remark on several properties of the conductivity. This expression is derived from Hassam's fluid equations. Similar expressions can be derived from kinetic electron treatment using specific collision models~\cite{drake1977, zocco2015}. In the collisional limit, where $k_{\parallel}^{2} D_{\parallel}/\omega \ll 1$, the conductivity becomes spatially uniform. As the collision effects become weaker, the parallel current and conductivity tend to peak around the singular point where $\bm{k}\cdot\bm{B} \approx 0$, similar to those obtained from kinetic theory. However, expression \eqref{eq:generalized_sigma} does not reproduce the collisionless expression because it is obtained under the assumption that $\omega/\nu_{\ele\ion} \ll 1$.

By substituting $\ept = -(\p^{2}/\p x^2 -k_{y}^{2})\apt/(\mu_{0}\sigma_{\para})$ into \eqref{eq:faraday_tearing}, we obtain
\begin{equation}
    \frac{\p \apt}{\p t} = \frac{1}{\mu_{0} \sigma_{\para}}
        \left( \frac{\p^{2}}{\p x^{2}} - k_{y}^{2} \right)
            \apt - \img k_{\para} \tilde{\phi}.
\label{eq:faraday_ohm_tearing}
\end{equation}
The second-order ordinary differential equations for $\tilde{\phi}$ \eqref{eq:vorticity_tearing} and $\tilde{A}_{\parallel}$ \eqref{eq:faraday_ohm_tearing} are solved using a standard multi-scale analysis~\cite{fkr1963}.

First, we consider the case without thermal forces and inertia ($\alpha_{\mathrm{t}},\alpha_{\mathrm{tt}}, \alpha_{\mathrm{i}}=0$). The parallel conductivity becomes
\begin{align}
    \sigma_{\para} = \frac{1}{\eta} \left(1 - \frac{\omegan}{\omega} - \frac{\omegat}{\omega} \right).
\end{align}
In the region around $B_{y}^{\mathrm{eq}} = 0$ (the inner region), $\p^2_x\gg k_y^2$ and $\by \approx B_{y0}(x/a)$ can be assumed. Subsequently, the eigenvalue equations are simplified to
\begin{align}
    - \img \omega \apt
    & =\left(1-\frac{\omegan}{\omega} - \frac{\omegat}{\omega} \right)^{-1} \frac{a^2}{\tauR}\frac{\p^2}{\p x^2} \apt+ B_{y0} \frac{x}{a}\vx,
\label{eq:eig1in}
\\
    \frac{\p}{\p t}\frac{\p^2}{\p x^2}\vx
    & =
    -\frac{1}{\tauA^2 B_{y0}} \frac{x}{a} \frac{\p^2}{\p x^2}\apt,
\label{eq:eig2in}
\end{align}
where $\tauA=1/(k_yV_{\mathrm{A}})$ and $\tauR=a^2\mu_0/\eta$ are the Alfv\'{e}n time and resistive time, respectively. Even though the equations can be solved analytically, we derive a scaling relation in a rather intuitive manner, implicitly employing knowledge of the analytic solution. By integrating \eqref{eq:eig1in} over the inner region ($0<x<\delta)$, we obtain
\begin{align}
    \left(1-\frac{\omegan}{\omega} - \frac{\omegat}{\omega} \right)^{-1} \frac{a^{2}}{\tauR} \int_{0}^{\delta} \frac{\p^{2}\apt}{\p x^{2}} \diff x
    = \int_{0}^{\delta}
        \left( - \img \omega \apt - B_{y0} \frac{x}{a} \vx \right) \diff x
    \RN{\propto
    -\img \omega \apt(0) \delta} .
\end{align}
The main part of the integral comes from the $\apt$ term, which is approximately constant and is denoted by $\apt(0)$ (the so-called {\it constant}--$A_{\parallel}$ approximation~\cite{fkr1963}). 
\RN{Since the exact solution of $\vx$ smoothly and asymptotically approaches to a constant of order unity away from the singular point~\cite{rutherford1971}}, the effect of the flow term in the integral is minor and is expressed by a numerical factor. (\RN{In this paper}, we exclusively \RN{set the proportionality factor as unity,} because this does not matter for scaling laws.) Using the definition of the tearing parameter $\Delta'$,
\begin{equation}
    \frac{\Delta'}{2} = \frac{1}{\tilde{A}_{\parallel}(0)} \left[ \frac{\p \tilde{A}_{\parallel}}{\p x} \right]_{0}^{\delta},
\end{equation}
we obtain
\begin{equation}
    - \img (\omega - \omega_{\ast \ele} ) =
    \frac{1}{\tau_{\mathrm{R}}} \frac{\Delta'a^2}{2\delta},
\label{eq:dtm_dipersion_relation}
\end{equation}
where we define $\omega_{\ast\ele}\equiv \omegan+\omegat$. The inner-layer width $\delta$ is estimated as $(\delta/a)^4 \approx \omega^2 \tau_{\mathrm{A}}^2/(\img (\omega - \omega_{\ast \ele}) \tau_{\mathrm{R}})$ from \eqref{eq:eig1in} and \eqref{eq:eig2in}. The integration is truncated at $\delta$ because the electric field (integrand) disappears in the outer region. Therefore, around $x \gg \delta$, where $\tilde{A}_{\parallel} \approx (\Delta'/2) \tilde{A}_{\parallel}(0) x$, the flow $\tilde{v}_{x}$ has the form
\begin{equation}
    \tilde{v}_{x} \approx - \frac{1}{B_{y0}} \frac{\Delta'a}{2} \img \omega \tilde{A}_{\parallel}(0).
\end{equation}
For a positive $\Delta'$, the flow points toward the X point.

\begin{figure}[ht]
\begin{center}
	\includegraphics[width=120mm]{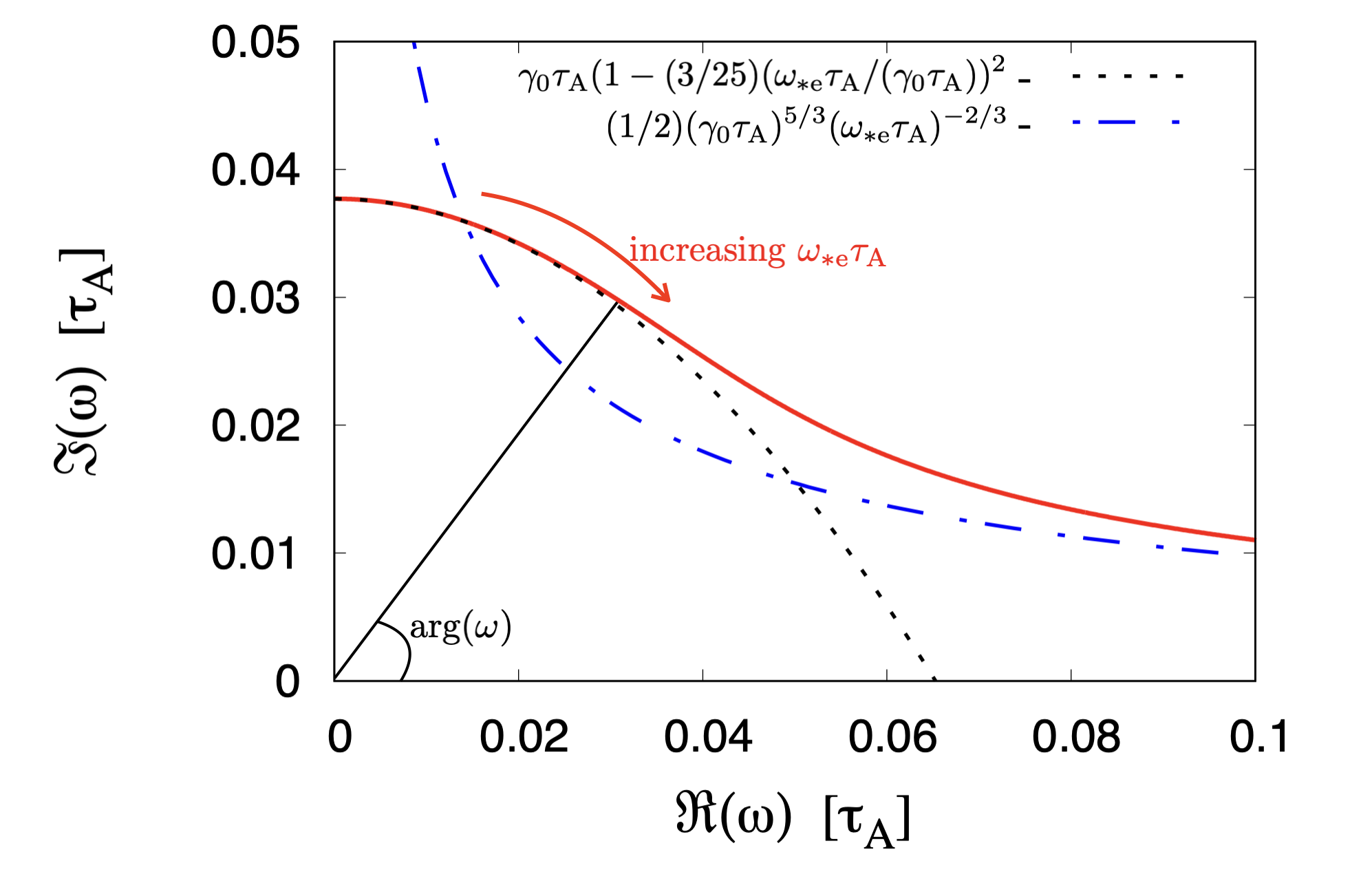}
\end{center}
	\caption{Numerical solution to \eqref{eq:dtm_dipersion_relation} in the complex $\omega$ plane 
	for $\omega_{\ast \ele} \tauA \geq 0$. The fixed parameter is $\gamma_0 \tauA = 0.038$ with $\Delta'a = 1$ and $S = 0.011$. The red line corresponds to the numerical solution, and the black and blue dashed lines represent the scaling for $\omega_{\ast \ele} \ll \gamma_0$ and $\omega_{\ast \ele} \gg \gamma_0$, respectively. The normal tearing mode, $\omega_{\ast\ele}\tauA=0$, is stabilized by the diamagnetic drift frequency $\omega_{\ast \ele}$ owing to the phase difference between $\vx$ and $\apt$.}
\label{fig:dtm_omega_plane}
\clearpage
\end{figure}

\begin{figure}[ht]
\begin{center}
	\includegraphics[width=160mm]{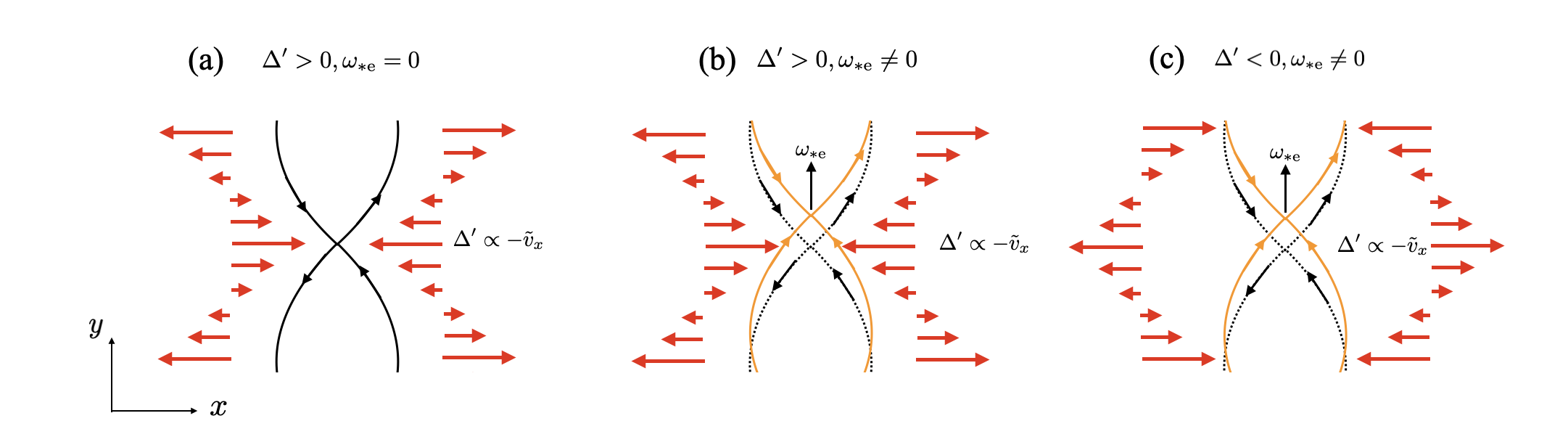}
\end{center}
	\caption{Schematic picture of phase relation between the magnetic and flow perturbations illustrating the stabilization of the normal tearing mode owing to the diamagnetic drift. The black dashed and solid lines correspond to the magnetic filed forming the X-point. The yellow lines represent the magnetic field drifting with the order of the electron diamagnetic drift frequency. The red arrows represent the $\vx$ direction depending on the sign of $\Delta '$.}
\label{fig:stabilizing_drift-tearing}
\clearpage
\end{figure}

If there is no drift, $\omega_{\ast\ele}=0$, we recover the normal tearing mode solution: $\gamma_{0}\tau_{\mathrm{A}} = \Im (\omega \tau_{\mathrm{A}})= (\Delta' a/2)^{4/5} S^{-3/5}$, where $S\equiv\tauR/\tauA$ is the Lundquist number. 
Solving \eqref{eq:dtm_dipersion_relation} for the limit $\omega_{\ast \ele} \ll \gamma_{0}$ or $\omega_{\ast \ele} \gg \gamma_{0}$, we obtain the drift-tearing mode solutions
\begin{align}
    \omega \tau_{\mathrm{A}}
        =
        \left\{
        \begin{matrix}
        \frac{3}{5} \omega_{\ast \ele}\tauA + \img \gamma_{0}\tauA
        \left(1 - \frac{3}{25} (\omega_{\ast \ele}/\gamma_{0})^{2} \right) &
            (\omega_{\ast \ele} \ll \gamma_{0}) \\
          \omega_{\ast \ele} \tau_{\mathrm{A}}
            + \img^{1/3} (\gamma_{0} \tauA)^{5/3} (\omega_{\ast \ele} \tauA)^{-2/3}
            & (\omega_{\ast \ele} \gg \gamma_{0})
        \end{matrix}
        \right..
\label{eq:dt_mode_freq}
\end{align}
In Fig.~\ref{fig:dtm_omega_plane}, a numerical solution to the dispersion relation \eqref{eq:dtm_dipersion_relation} is shown in the complex $\omega$ plane. As $\omega_{\ast\ele}$ increases, the growth rate decreases and asymptotes to scaling $\omega_{\ast\ele}^{-2/3}$. The argument of $\omega$ decreases from $\pi/2$ to zero, which corresponds to the phase difference between $\tilde{A}_{\parallel}$ and $\tilde{v}_{x}$ (see \eqref{eq:eig2in}). Figure~\ref{fig:stabilizing_drift-tearing} shows how the drift stabilizes the tearing mode. In Fig.~\ref{fig:stabilizing_drift-tearing} (a), a normal tearing case is shown. The black solid lines represent the magnetic field lines, and the red arrows indicate the direction of the plasma flows. The direction of the flow corresponding to the sign of $\Delta'$ is determined by the competition between the magnetic tension force and pressure force. Even though the magnetic tension force tends to expel plasmas outward (opposite direction to the red arrows), plasmas are attracted to the X point to maintain the incompressibility. Therefore, the flux is continuously supplied and grows exponentially. When finite drift is incorporated, the amount of flux supplied to the X point is reduced because the location of the maximum flow slips off from the X point (Fig.~\ref{fig:stabilizing_drift-tearing}(b)). In the negative $\Delta'$ case (Fig.~\ref{fig:stabilizing_drift-tearing}(c)), the flows are in opposite directions. Therefore, the magnetic perturbation is damped. One might expect that the drift destabilizes the mode for $\Delta'<0$ if it convects the X point to the position where the flow points inward. However, this never occurs because the phase difference never exceeds zero.

We now discuss the destabilization mechanism of collisional microtearing instability ($\at,\att,\ai \neq 0$). The parallel conductivity becomes
\begin{align}
    \sigma_{\parallel} 
    = \frac{1}{\eta}
    \left[
        1-\frac{\omegan}{\omega}-\frac{\omegat}{\omega}\left(1+\at+ \attt \frac{\img\omega}{\nu_{\ele\ion}} \right)
    \right] 
    \left(
        1-\frac{\ai}{\alpha_{\eta}}\frac{\img \omega}{\nu_{\ele\ion}}
    \right)^{-1}.
\label{eq:generalized_sigma_mtm}
\end{align}
By substituting \eqref{eq:generalized_sigma_mtm} into \eqref{eq:faraday_ohm_tearing}, and integrating over the inner layer $\delta_{\mathrm{MT}}$, we obtain
\begin{align}
    -\img \omega \left(1-\frac{(\omegan+(1+\at) \omegat)}{\omega} - \attt \frac{\img \omegat} {\nu_{\ele\ion}}\right)
    \left(1 - \frac{\ai}{\alpha_{\eta}}\frac{\img \omega}{\nu_{\ele\ion}} \right)^{-1}
    =
    \frac{1}{c} \frac{1}{\tauR} \frac{\Delta'a}{2\delta_{\mathrm{MT}}}.
\label{eq:mt_apt}
\end{align}
The layer width $\delta_{\mathrm{MT}}$ is given by
\begin{align}
    \frac{\delta_{\mathrm{MT}}^4 }{a^4}
    =\frac{\tauA^2}{\tauR}\frac{\omega^2 }{\img(\omega-(\omegan+(1+\at)\omegat) 
    -\attt\omegat (\img \omega/\nu_{\ele\ion}))}\left( 1- \frac{\ai}{\alpha_{\eta}} \frac{\img \omega}{\nu_{\ele\ion}}\right).
\label{eq:delta_mt}
\end{align}
By substituting \eqref{eq:delta_mt} into \eqref{eq:mt_apt}, and using the collisional condition $\omega/\nu_{\ele\ion}\ll 1$, we obtain the following dispersion relation for the collisional microtearing mode:
\begin{align}
    \left[
        1 - \frac{\omega_{\ast\mathrm{MT}}}{\omega} - \img\frac{\omegat}{\nu_{\ele\ion}}
            \left(
                \attt+\frac{\ai}{\alpha_{\eta}}\frac{\omega_{\ast\mathrm{MT}}}{\omegat}
            \right)
    \right]^3
    =
    \frac{\img}{\omega^5} \left( \frac{\Delta'a}{2 c\tauR^{3/4} \tauA^{1/2}}\right)^4
    = \img \left(\frac{\gamma_{0}}{\omega}\right)^5,
\label{eq:mt_disp}
\end{align}
where $\omega_{\ast\mathrm{MT}} \equiv \omegan+(1+\at) \omegat$. For simplicity, we define a new parameter $\alpha_{\mathrm{tt,i}}\equiv \attt +(\ai/\alpha_{\eta})(\omega_{\ast\mathrm{MT}}/\omegat)$. We look for a solution
\begin{align}
    \frac{\omega}{\nu_{\ele\ion}} \sim
    \frac{\omega_{\ast n_{0\ele}}}{\nu_{\ele\ion}} \sim
    \frac{\omega_{\ast T_{0\ele}}}{\nu_{\ele\ion}} \sim
    \epsilon \ll 1.
\end{align}
If $(\gamma_{0}/\omega)^{5/3} \sim O(1)$, we arrive at the same solution as in the drift-tearing case, which is stable if $\Delta'<0$. However, if we consider $(\gamma_{0}/\omega)^{5/3}\sim \epsilon$, we obtain the following unstable microtearing mode solution:
\begin{align}
    \omega\tauA
        =
          \omega_{\ast \mathrm{MT}} \tauA \left(1 + \alpha_{\mathrm{tt,i}}
          \frac{\img\omega_{\ast T_{0\ele}}}{\nu_{\ele\ion}} \right)
        + \img^{1/3} (\gamma_{0}\tauA)^{5/3}(\omega_{\ast\mathrm{MT}}\tauA)^{-2/3}.
\label{eq:mt_mode_freq}
\end{align}
If $\gamma_{0}/\omega$ is even smaller, we recover the same dispersion relation as in the self-filamentation mode.
Because the second term including $\gamma_{0}$ stabilizes the microtearing mode if $\Delta'<0$, only the term proportional to $\alpha_{\mathrm{tt,i}}$ causes instability. The destabilization mechanism is the same as in the self-filamentation mode, that is, the lag of the inductive electric field behind the magnetic field owing to the time-dependent thermal force and electron inertia. Note that these two effects result in essentially the same effect and are indistinguishable. Hereafter, we refer to this force as time-dependent thermal/inertia force. 

As discussed in~\cite{drake1983}, the spatial structure of the microtearing mode is significantly different from that of the normal and drift-tearing modes, which are driven by the macroscopic magnetic field characterized by $\Delta'$. When the thermal conductivity is invoked in the weakly collisional case, the temperature fluctuations are washed out at around $\Delta_{\kappa}$, where $\Delta_{\kappa}^{2}\equiv\omega/({k_{\parallel}'}^2 \kappa_{\parallel})$. Consequently, the plasma attempts to move faster near $k_{\parallel}=0$. The resultant diamagnetic current near the singular point shields the magnetic perturbation in the outer region from the singular point. If magnetic diffusion over a single wave period $\sim\omega_{\ast\ele}$ does not reach $\Delta_{\kappa}$, such that
\begin{equation}
    \Delta_{\kappa} \gg \Delta_{\eta} \equiv \left(\frac{\eta}{\mu_{0}\omega_{\ast T_{0\ele}}} \right)^{1/2},
\end{equation}
the mode is localized near the singular point. The localization is characterized by $\hat{\beta}_{\mathrm{T}}\equiv (\beta_{\ele}/2) (B_{z0} a/(B_{y0} L_{T_{0\ele}}))^{2} \propto \Delta_{\kappa}^{2}/\Delta_{\eta}^{2}$. For large $\hat{\beta}_{\mathrm{T}}$, the microtearing mode is disconnected from the outer region and thus is independent of $\Delta'$. From the scaling, the current layer width of the microtearing mode in the leading order is scaled as $\delta_{\mathrm{MT}}/a \sim (a/(Sd_{\ele}))^{1/2}$. (For the normal tearing mode, $\delta/a \sim (\gamma_{0}\tauA/S)^{1/4}$.)

Finally, we comment on the collisionless case, although the analyses presented here are not applicable to the collisionless case because they are based on the collisional model. As the collision frequency decreases, the destabilization term may increase as long as $\alpha_{\mathrm{tt,i}}$ is constant. However, as might be expected, the effects of the thermal forces and inertia induced by collisions must vanish in the collisionless limit, that is, $\alpha_{\mathrm{tt,i}} \rightarrow 0$. From the numerical simulations, we observe that the microtearing mode is stabilized in a weakly collisional regime. 
In the collisionless regime, the {\it{genuine}} electron inertia works \MY{as an effective magnetic diffusion within $d_{\ele}$} near the singular point. 
\MY{However, the magnetic shielding also works to stabilize the mode if the condition $(\Delta_{\kappa}/d_{\ele})^2 \gg 1$ is satisfied, which is read as
\begin{equation}
    \hat{\beta}_{\mathrm{T}} \gg \frac{2\alpha_{\kappa}}{3}
    \left(\frac{\eta_{\ele}}{\eta_{\ele}+1}\right)^2
    \frac{\omega_{\ast \ele}}{\nu_{\ele\ion}}.
\label{eq:stabilization_condition_collisionless_mt}
\end{equation}}
We speculate that \MY{as long as the condition \eqref{eq:stabilization_condition_collisionless_mt} is met,} the {\it{genuine}} inertia does not contribute to the collisionless microtearing mode destabilization, \MY{as shown by Drake and Lee~\cite{drake1977}}.


\section{Gyrokinetic simulation analysis of the microtearing mode}
\label{sec:simulation}

\subsection{Simulation setup}
\label{subsec:setup}
We perform linear and nonlinear gyrokinetic simulations of the microtearing mode using $\agk$ code. The simulation setup follows that of a previous tearing instability study~\cite{numata2011}, which is equivalent to the configuration discussed in Sec.~\ref{subsec:tearing_analysis}. For the microtearing instability study, we stabilize the normal tearing mode by setting $\Delta'<0$ and additionally apply the electron temperature gradient $L_{T_{0\ele}}$ and density gradient $L_{n_{0\ele}}$.

The basic parameters are chosen as follows: $\beta_{\ele}=0.024$ with reference to NSTX~\cite{moradi2013}, $k_ya=2.5$ ($\Delta ^{\prime}a =-4.58$), $B_{z0}a/(B_{y0}L_{T_{0\ele}})=500$, $\eta_{\ele}\equiv L_{T_{0\ele}}^{-1}/L_{n_{0\ele}}^{-1}=1$,
$n_{0\ele}=n_{0\ion}=n_{0}$, $\tau \equiv T_{0\ion}/T_{0\ele}=1$, $\sigma \equiv m_{\ele}/m_{\ion}=1/1836$, $Z=-q_{\ion}/e=1$, and $\rho_{\mathrm{Se}}/a=0.0141$, where $\rho_{\mathrm{Se}} \equiv \sqrt{m_{\ion}T_{0\ele}}/(eB_{z0})$ is the ion-sound Larmor radius. We scan the collision frequency $\nu_{\ele\ion}$ ranging from $\nu_{\ele\ion}/\omega_{\ast\ele}\approx100$ (highly collisional regime) to $\nu_{\ele\ion}/\omega_{\ast\ele}\approx0.01$ (weakly collisional regime).

As discussed in Sec.~\ref{subsec:tearing_analysis}, the destabilization physics of the slab microtearing mode is the time-dependent thermal/inertia force (see \eqref{eq:mt_mode_freq}). To highlight this destabilization mechanism, we choose parameters that satisfy $\omega_{\ast \mathrm{MT}} \tauA \sim \omega_{\ast\ele} \tauA \gg (\gamma_0 \tauA)^{5/2}$, where $ \omega_{\ast\ele} \tauA$ can be represented by $\hat{\beta}_{\mathrm{T}}$ as $ \omega_{\ast\ele} \tauA = (\rho_{\mathrm{Se}}/a) (\hat{\beta}_{\mathrm{T}})^{1/2}(1/\eta_{\ele} + 1)$. For the parameters given above, $6.48 \times 10^{-11} \lesssim (\gamma_{0}\tauA)^{5/2} \lesssim 6.48 \times 10^{-5}$ for $0.01\leq\nu_{\ele\ion}/\omega_{\ast\ele}\leq 100$ and $\omega_{\ast\ele}\tauA=1.55$ ($\hat{\beta}_{\mathrm{T}}=3000$); therefore, $\omega_{\ast\ele} \tauA \gg (\gamma_0 \tauA)^{5/2}$ is well satisfied.

Collision physics is a key component of this study.
$\agk$ employs the model Fokker--Planck collision operator, which includes pitch-angle scattering, energy diffusion, and conserving terms to ensure the conservation of the particle number, momentum, and energy. Basically, the collisionality is dependent on the particle velocity; therefore, the velocity-dependent collision frequencies for various collision operators are given by the collision frequency $\nu_{st}$ between the particle species $s$ and $t$, multiplied by some function of $v/v_{\mathrm{th}}$. For example, the Lorentz collision operator includes $\nu_{st}(v/v_{\mathrm{th},s})^{-3}$. Such velocity dependences in the collision operators can be switched off artificially in \agk. We show that these velocity-dependent collision frequencies are necessary for the microtearing mode to become unstable, as theoretically predicted. In this study, we consider the electron--electron and electron--ion collision frequencies as free parameters and set $\nu_{\ele\ele}=\nu_{\ele\ion}$. We ignore the ion collisions $\nu_{\ion\ion}=\nu_{\ion\ele}=0$.
Unless otherwise stated, we use the full collision operator including pitch-angle scattering, energy diffusion, and conserving terms with the velocity-dependent collision frequency.

\subsection{Linear simulation results}
\label{sec:simulation_results}
\subsubsection{Properties of the linear microtearing mode}

\begin{figure}[htbp]
	\begin{minipage}{0.5\hsize}
		\begin{center}
		\includegraphics[width=90mm]{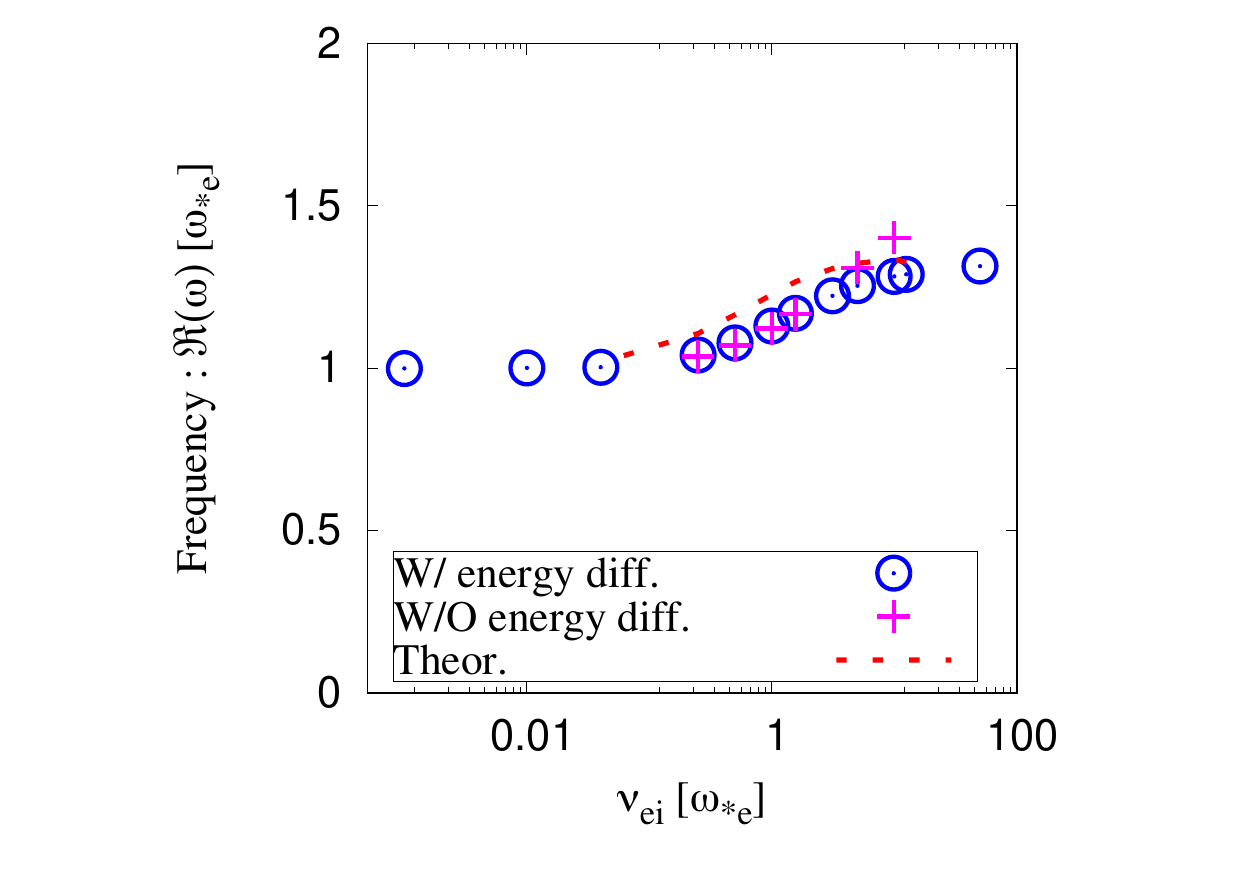}
		\end{center}
	\end{minipage}
	\begin{minipage}{0.5\hsize}
		\begin{center}
		\includegraphics[width=90mm]{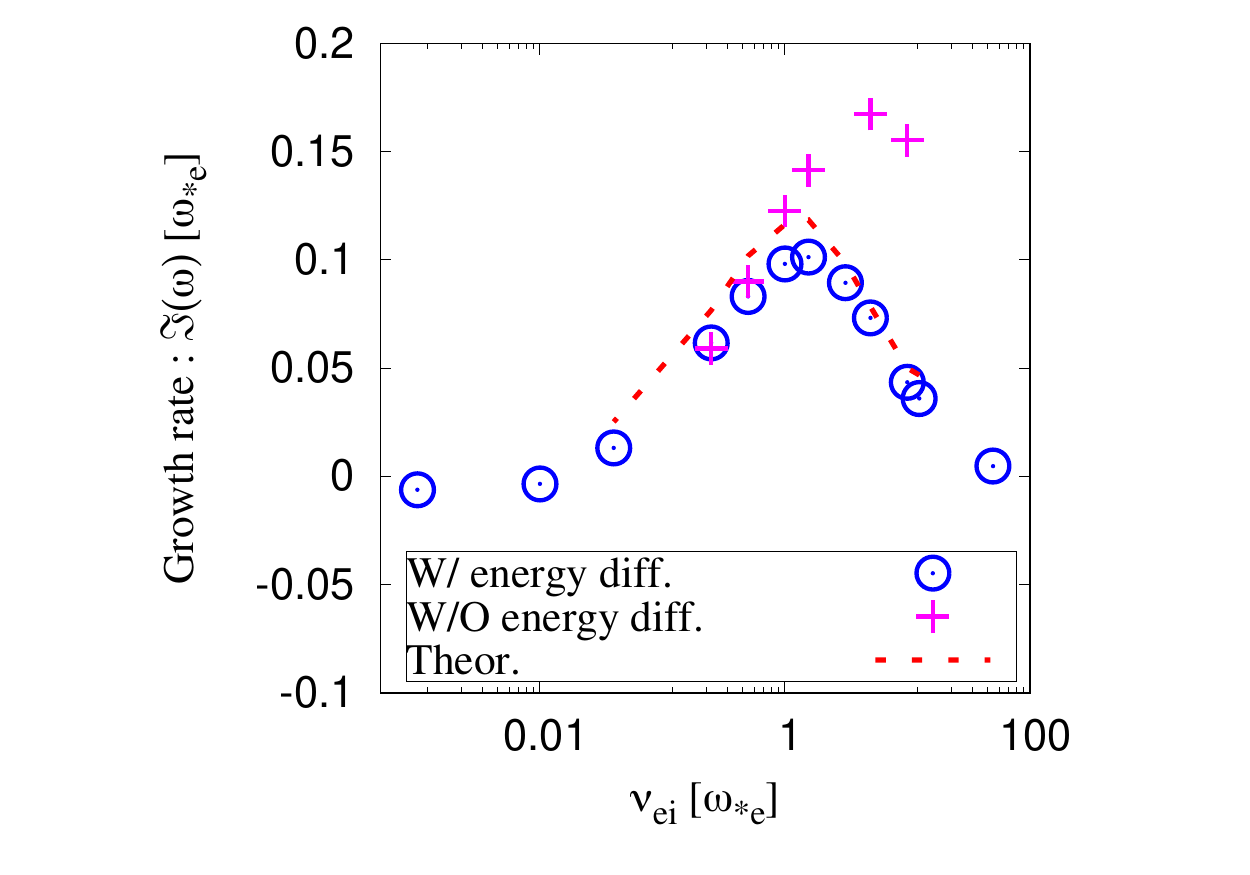}
		\end{center}
	\end{minipage}
	\caption{Collision-frequency dependence of the microtearing mode eigen-frequencies. The eigen-frequencies and collision frequency are normalized by the electron diamagnetic frequency $\omega_{\ast\ele} =\omega_{\ast T_{0\ele}}+\omega_{\ast n_{0\ele}}$. The blue circles and magenta crosses correspond to cases with and without the energy diffusion collision operator, respectively. The red dashed line represents the theoretical curve corresponding to \eqref{eq:mt_apt} obtained using the coefficient $\at$ and $\atti$ measured from the simulation results, as shown later in Fig.~\ref{fig:coeff}. The fixed parameters are $\beta_{\ele}=0.024$, $\Delta ^{\prime}a=-4.58$, $B_{z0}a/(B_{y0}L_{T_{0\ele}})=500$, $\eta_{\ele}=1$, $\tau=1$, $\sigma=1/1836$, and $\rho_{\mathrm{Se}}/a = 0.0141$. For these parameters, $\omega_{\ast \ele} \tauA = 1.55$ and $\hat{\beta}_{\mathrm{T}}=3000$. The real frequency is $\omegar\sim \omega_{\ast\ele}$ in all collision regimes. The growth rate peaks at $\nu_{\ele\ion} \sim \omega_{\ast \ele}$ and becomes negative both in weakly or strongly collisional regimes. }
\label{fig:nuscan}
\end{figure}

We solve the linear initial value problem and obtain the complex eigen-frequencies for the most unstable (least stable) mode with the given parameters.
Figure~\ref{fig:nuscan} shows the collision-frequency dependence of the microtearing mode.
The collision frequency $\nu_{\ele\ion}$, mode growth rate $\omegai$, and real frequency $\omegar$ are normalized by the diamagnetic drift frequency $\omega_{\ast\ele} = \omega_{\ast T_{0\ele}}+\omega_{\ast n_{0\ele}}$. The blue circles and magenta crosses correspond to cases with and without the energy diffusion operator, respectively. The red dashed line is the theoretical curve corresponding to \eqref{eq:mt_apt} obtained using the coefficients $\at$ and $\atti$ measured from the simulation results, as shown later in Fig.~\ref{fig:coeff}.
The real frequencies in the \RN{left} figure are $\omegar \sim \omega_{\ast \ele}$ for all regimes from $\nu_{\ele\ion}/\omega_{\ast\ele}\approx 100$ to $\nu_{\ele\ion}/\omega_{\ast\ele}\approx 0.01$ and are almost independent of the collision operators. The growth rates in the \RN{right} figure show a non-monotonic dependence on the collision frequency for both collision operators. The growth rate peaks at $\nu_{\ele\ion}/\omega_{\ast_\ele}=1.56$ for the collision operator with energy diffusion, whereas it peaks at $\nu_{\ele\ion}/\omega_{\ast_\ele}=5.0$ for the operator without energy diffusion, which is approximately 1.6 times higher than the maximum growth rate obtained using energy diffusion. 
The energy diffusion operator strongly suppresses the mode growth. This discrepancy in collision operators is most remarkable in $1\lesssim \nu_{\ele\ion}/\omega_{\ast\ele}\lesssim10$.
It should be noted that linear stability is quite sensitive to the collision model, particularly in the semi-collisional regime.
The microtearing mode is stable at both collisionless and collisional limits. These results qualitatively agree with those of previous studies~\cite{dippolito1980, gladd1980}. The stability threshold is reached at approximately $\nu_{\ele\ion}/\omega_{\ast\ele}\sim 0.01$. 
The microtearing mode may also exist in weakly collisional plasmas in future high-temperature plasma confinement devices.
\MY{The} applicability of our result to the present and future experiments is discussed in Sec.~\ref{sec:summary}. 

\begin{figure}[htbp]
	\begin{center}
	\includegraphics[width=165mm]{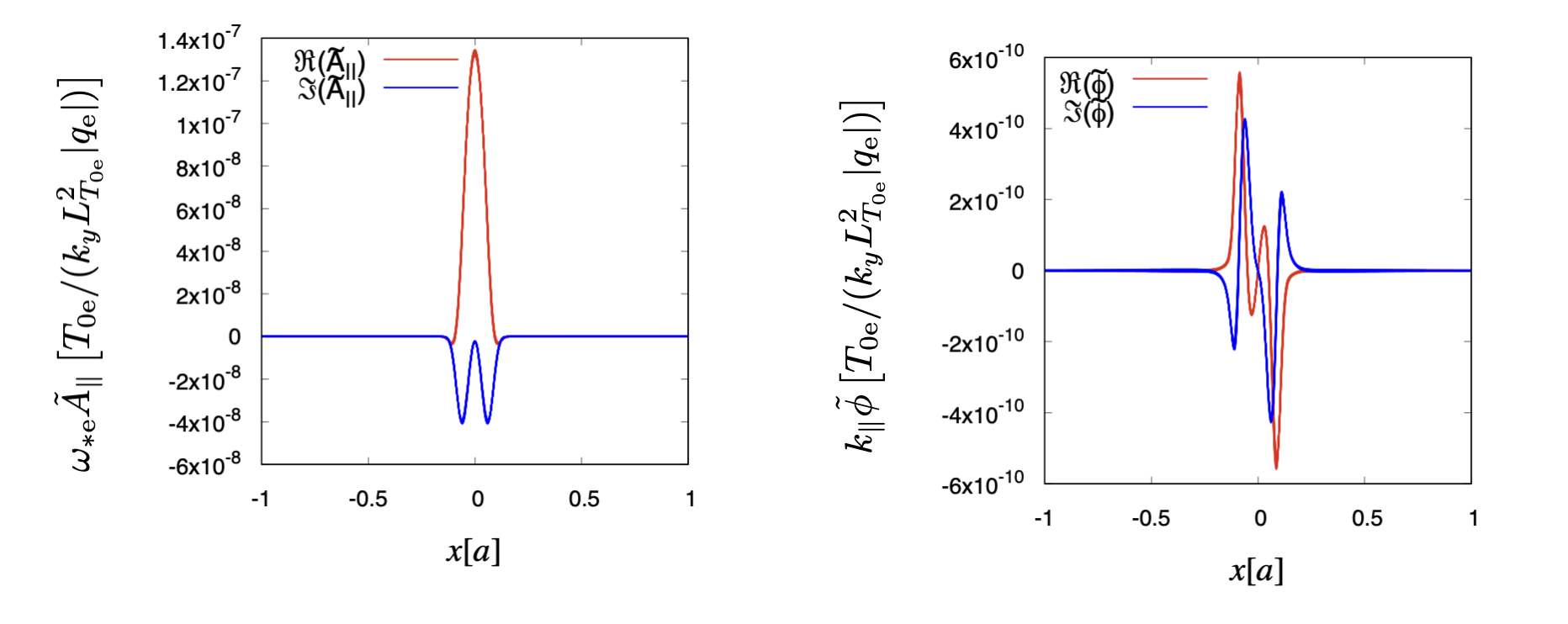}
	\end{center}
	\caption{Eigen-functions of the microtearing mode at $\nu_{\ele\ion}/\omega_{*e}=1.56$. The parallel magnetic potential $\tilde{A}_{\para}$ (left) and electrostatic potential $\tilde{\phi}$ (right) represented in the dimensions of the parallel electric field are shown. The mode structures are tearing parity. The red and blue lines are real and imaginary parts, respectively. The amplitude of $\apt$ is significantly larger than that of $\tilde{\phi}$.}
\label{fig:eigf}
\clearpage
\end{figure}

Figure~\ref{fig:eigf} shows the typical eigen-functions of $\tilde{A}_{\para}$ and $\tilde{\phi}$ in microtearing mode. To compare the amplitudes of these functions, $\tilde{A}_{\para}$ and $\tilde{\phi}$ are represented in the dimensions of the electric field, that is, $\omega \tilde{A}_{\para} \sim \omega_{\ast \ele} \tilde{A}_{\para}$ and $k_{\para} \tilde{\phi}$, and normalized by $T_{0\ele}/(k_y L_{T_{0\ele}}^2 |q_{\ele}|)$.
The red and blue lines represent real and imaginary parts, respectively. $\apt$ is an even function, and $\tilde{\phi}$ is an odd function, that is, tearing parity. The mode structure is localized around the singular point $x/a=0$.
The magnetic perturbation is clearly dominant, and the electrostatic potential is minor.

\begin{figure}[htbp]
	\begin{minipage}{0.5\hsize}
		\begin{center}
		\includegraphics[width=85mm]{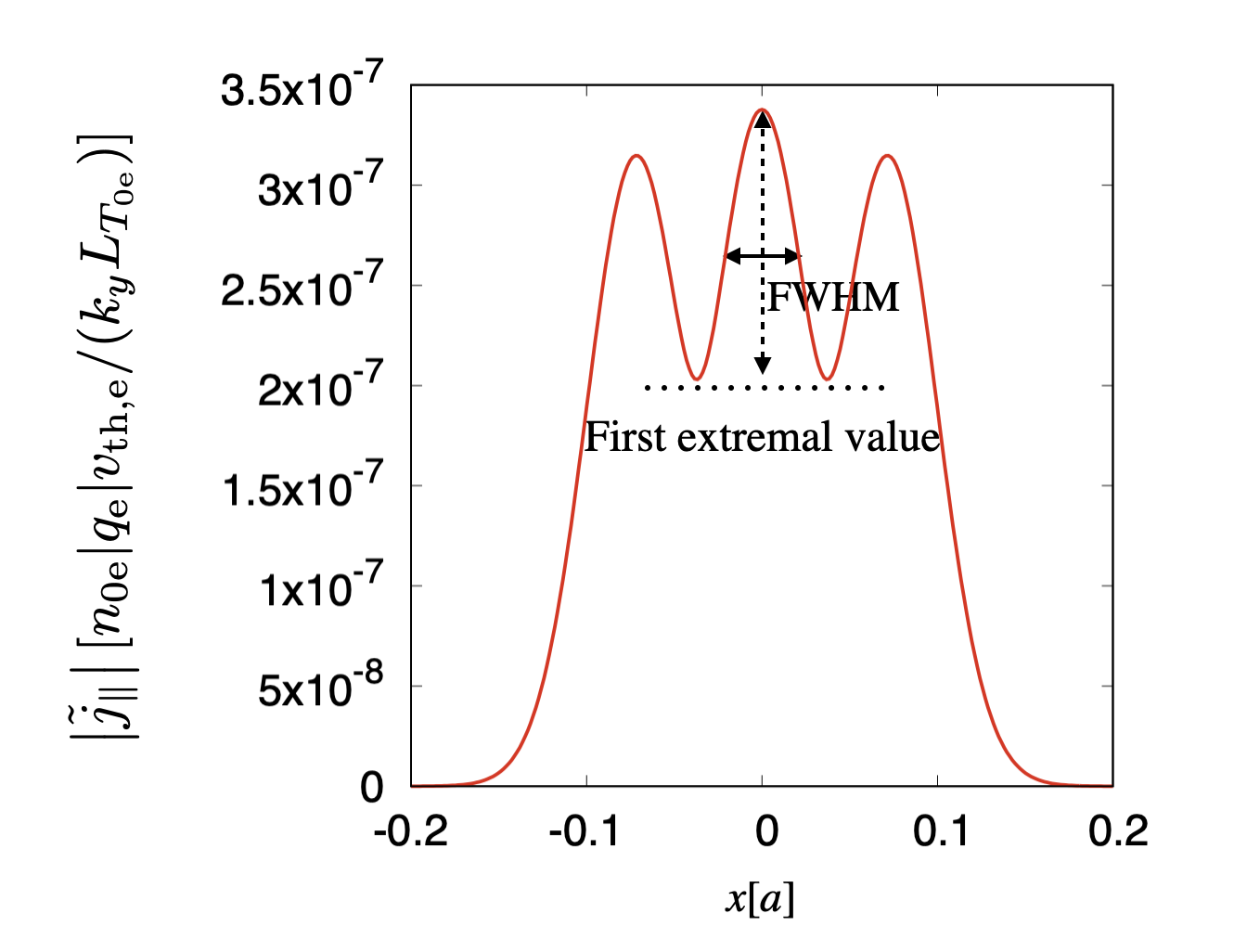}
		\end{center}
	\end{minipage}
	\begin{minipage}{0.5\hsize}
		\begin{center}
		\includegraphics[width=90mm]{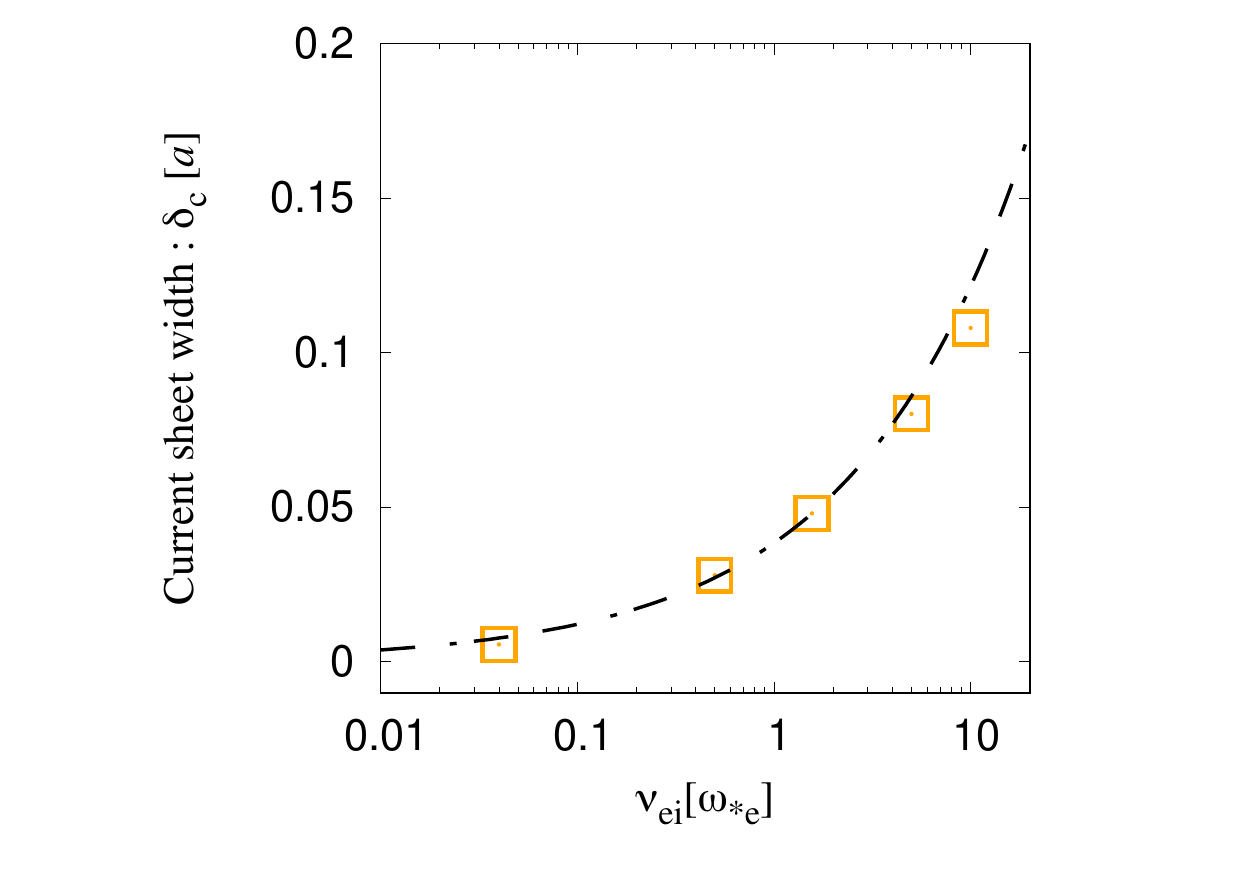}
		\end{center}
	\end{minipage}
	\caption{Definition of the current layer measurement (left) and the measured current layer widths (right). The current layer width $\delta_{\mathrm{c}}/a$ is defined by the full width at half maximum height of $|\tilde{j}_{\para}|$ where the height is measured from the top to the first local minimal value. In the right figure, the measured current layer widths follow the scaling $\delta_{\mathrm{MT}}/a \sim S^{-1/2}$.}
\label{fig:current_sheet}
\clearpage
\end{figure}

The current layer width is measured using the eigen-function $\tilde{j}_{\para}$ shown in Fig.~\ref{fig:current_sheet} (left). $\tilde{j}_{\para}$ is normalized by \RN{$n_{0\ele} |q_{\ele}| \vthe / (k_y L_{T_{0\ele}})$}. As shown in the figure, the current profile may exhibit multiple peaks. We define the current layer width $\delta_{\mathrm{c}}$ as the central peak of $|\tilde{j}_{\para}|$ such that $\delta_{\mathrm{c}}$ is given by the full width at half maximum height of $|\tilde{j}_{\para}|$ where the height is measured from the top to the first  local minimum value.
Figure~\ref{fig:current_sheet} (right) shows the current layer width $\delta_{\mathrm{c}}/a$ against $\nu_{\ele\ion}/\omega_{\ast\ele}$. The measured layer widths follow the scaling of $\delta_{\mathrm{MT}}/a \sim S^{-1/2}$ and represent theoretical predictions.


\begin{figure}[htp]
	\begin{minipage}{0.5\hsize}
		\begin{center}
		\includegraphics[width=90mm]{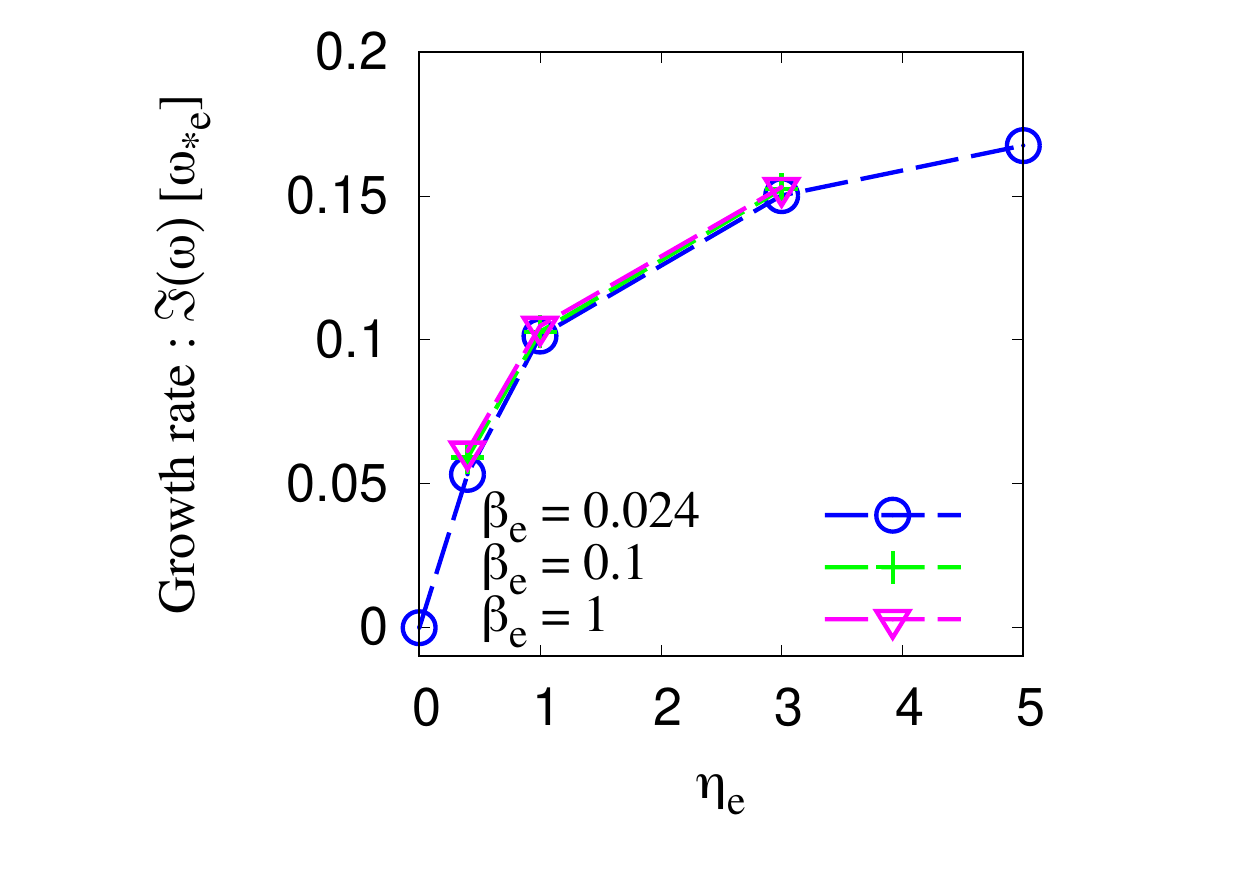}
		\end{center}
	\end{minipage}
	\begin{minipage}{0.5\hsize}
		\begin{center}
		\includegraphics[width=90mm]{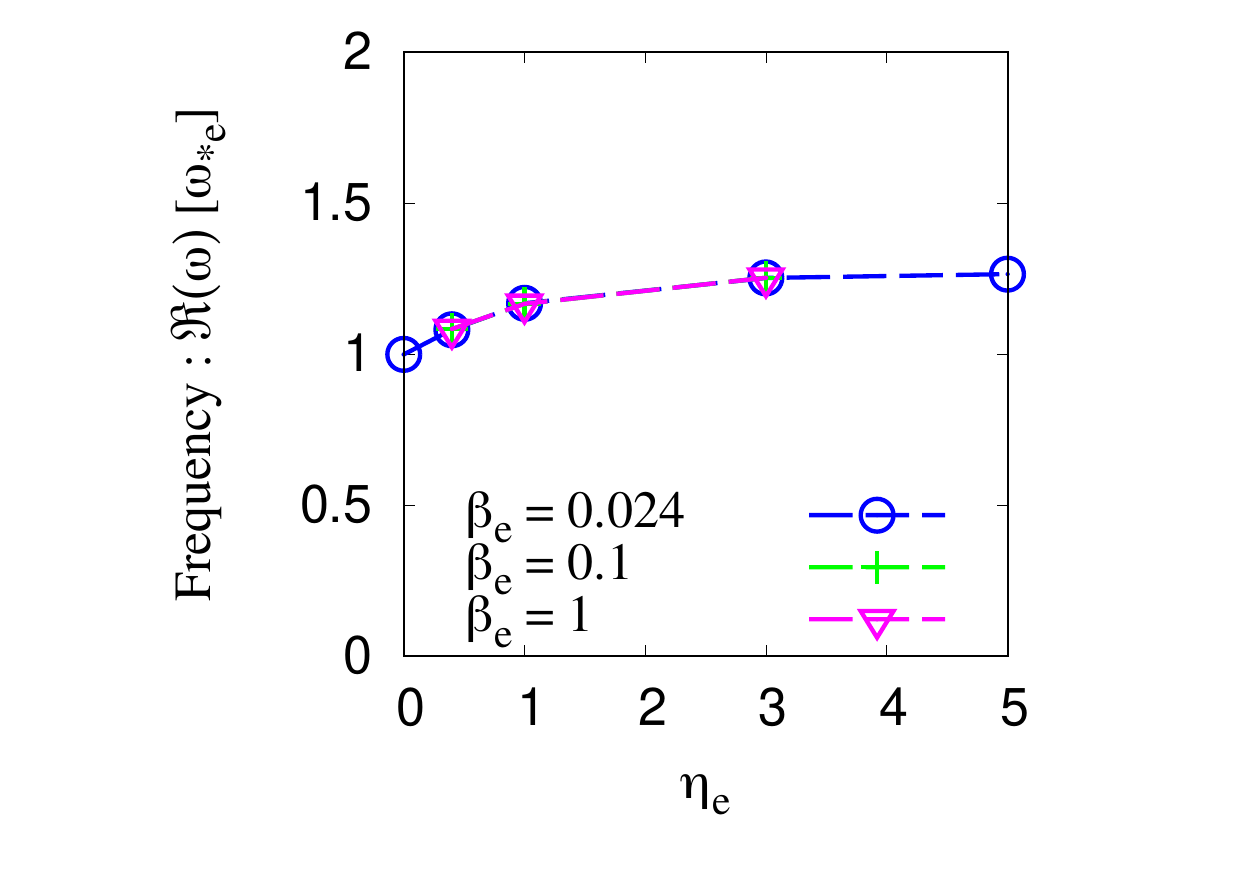}
		\end{center}
	\end{minipage}
	\caption{Dependence of the microtearing mode with respect to the electron temperature gradient $\eta_{\ele}$ for several $\beta_\ele$ values. The other parameters except the $\eta_\ele$ and $\beta_\ele$ are the same as those in the most unstable case, with energy diffusion shown in Fig.~\ref{fig:nuscan}, that is, $\nu_{\ele\ion}/\omega_{\ast_\ele}=1.56$. The normalized parameter of the mode eigen-frequencies is also the same as that in Fig.~\ref{fig:nuscan}.}
\label{fig:ltscan}
\clearpage
\end{figure}

Figure~\ref{fig:ltscan} shows the electron temperature gradient dependence of the microtearing mode eigen-frequencies for several electron beta values, $\beta_{\ele}=0.024, 0.1, 1$. The other parameters are the same as those in the most unstable case, with energy diffusion shown in Fig.~\ref{fig:nuscan} , that is, $\nu_{\ele\ion}/\omega_{\ast_\ele}=1.56$. 
The growth rates in the left figure monotonically increase with respect to $\eta_\ele$.
The mode is stable at $\eta_\ele=0$. The real frequencies in the right figure are always of the order of the diamagnetic drift frequency.
Clearly, the plasma beta has no effect on this parameter regime. It is predicted that the eigen-frequency is independent of $\beta_{\ele}$ provided that $\gamma_{0}/\omega_{\ast\ele}$ is small  \RN{because the time-dependent thermal force term is irrespective of $\beta_{\ele}$ (see \eqref{eq:mt_mode_freq}).}
We may expect that the level of microtearing turbulence in future higher-$\beta$ devices will be driven at the same level as that observed in the present tokamaks \RN{as long as} the assumption\RN{, $\gamma_{0}/\omega_{\ast\ele}\ll 1$,} is still holds.

Finally, we investigate the effect of the velocity-dependent collision frequency by artificially switching off the velocity dependence. We perform a simulation for the same parameter set as the most unstable case in Fig.~\ref{fig:nuscan}, but without the velocity dependence of the collision frequency, and obtain $\omega/\omega_{\ast\ele}=0.968 - \img 0.0108$.
This result implies that the effect not included in the velocity-independent collision, which is the time-dependent thermal/inertia force,  is the source of instability. 
In the next section, we show that the quantitative measurement of the coefficient $\atti$ of the time-dependent thermal/inertia force allows us to explain the microtearing unstable regime against collisionality.

\subsubsection{Evaluation of the time-dependent thermal force}
We estimate the coefficient $\atti$ of the time-dependent thermal/inertia force from the simulation results. At the singular point $k_{\parallel}(x)=0$, the parallel electric field is given solely by the inductive components, $\ept = \img \omega \apt = \tilde{j}_{\para}/\sigma_{\para}$. Using \eqref{eq:generalized_sigma_mtm} for $\sigma_{\para}$, we obtain the following equation:
\begin{align}
    \img \at \omegat \apt  -  \atti \frac{\omega}{\nu_{\ele\ion}} \omegat \apt
	=
    \img \omega \apt - ( \img (\omegan + \omegat) \apt + \eta \tilde{j}_{\para} ),
\label{eq:coefficient_measurement}
\end{align}
where the LHS represents the electric field generated by the thermal forces and collisional inertia force. We assume that the coefficient of resistivity $\alpha_{\eta}=0.380$ is fixed. By substituting the complex-valued $\apt$, $\tilde{j}_{\para}$, and $\omega$ obtained from the simulations in \eqref{eq:coefficient_measurement}, the coefficients $\at$ and $\atti$ are uniquely determined.

\begin{figure}[htbp]
\begin{center}
	\includegraphics[width=100mm]{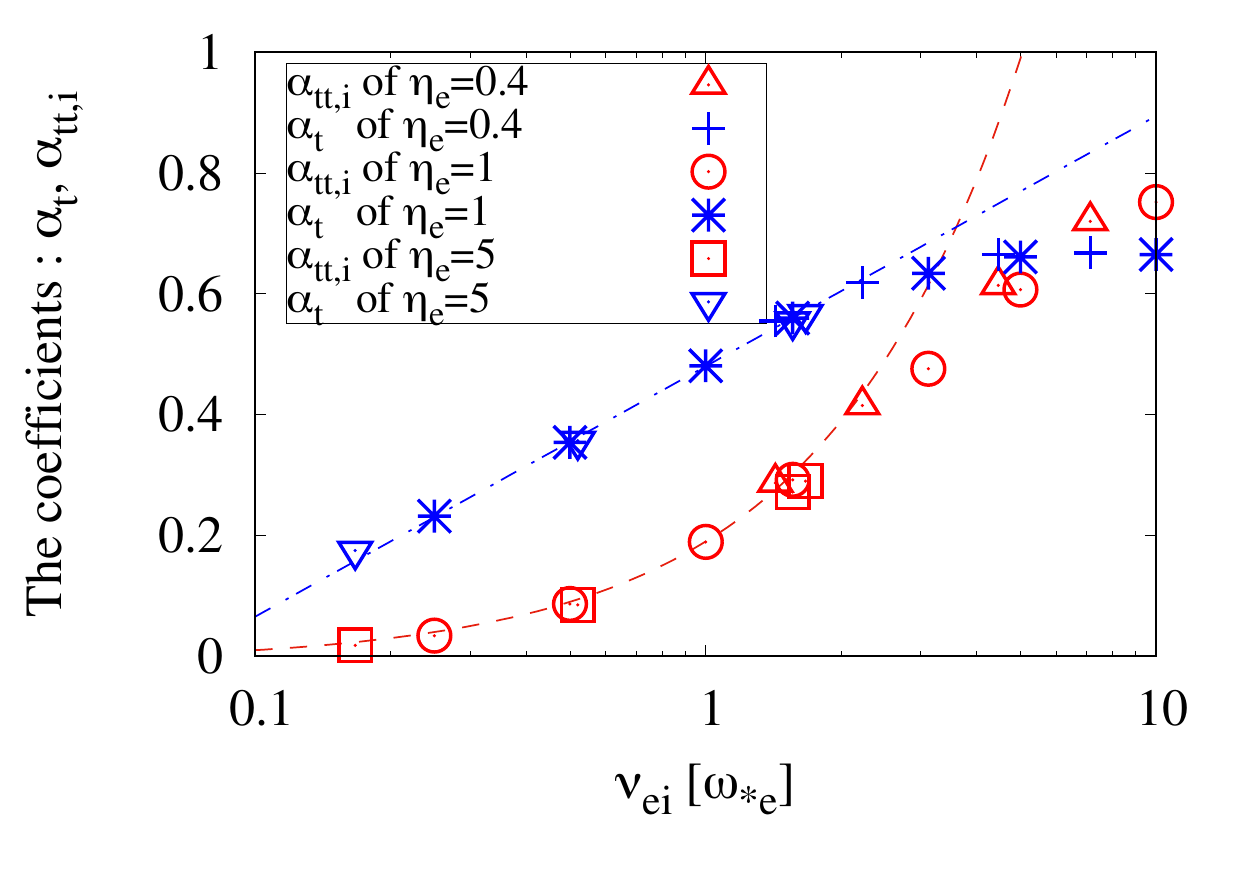}
\end{center}
	\caption{Coefficients of the thermal and time-dependent thermal/inertia forces, $\at$ and $\atti$, with respect to the collision frequency $\nu_{\ele\ion}$ normalized by $\omega_{\ast \ele}$ for the different electron temperature gradient $\eta_\ele$. The red and blue symbols correspond to $\atti$ and $\at$, respectively. Parameters other than $\eta_\ele$ are chosen to be the same as those in the most unstable case with energy diffusion in Fig.~\ref{fig:nuscan}. The coefficients $\atti$ and $\at$ depend on the collision frequency in $\nu_{\ele\ion}\lesssim \omega_{\ast \ele}$, vanish at $\nu_{\ele\ion}\ll\omega_{\ast\ele}$, and tend to become constant in $\nu_{\ele\ion}\gg \omega_{\ast \ele}$.}
\label{fig:coeff}
\end{figure}

Figure~\ref{fig:coeff} shows the measured coefficient $\atti$ of the time-dependent thermal/inertia force against the collision frequency $\nu_{\ele\ion}/\omega_{\ast \ele}$ for different electron temperature gradient $\eta_\ele$ values. 
The coefficient $\at$ of the thermal force is also plotted in Fig.~\ref{fig:coeff}. The red and blue symbols correspond to $\atti$ and $\at$, respectively. Parameters other than the electron temperature gradient are chosen to be the same as those in the most unstable case with energy diffusion in Fig.~\ref{fig:nuscan}. The coefficients $\at$ and $\atti$ should be constant with respect to the collision frequency in the collisional limit. However, the collision-frequency dependence is clearly observed in Fig. \ref{fig:coeff}. It is observed that, as the collision frequency increases, the coefficients tend to become constants. Referring to~\cite{hassam1980_1}, the coefficients of the Lorentz collision operator are $\at \approx 0.8$ and $\atti \approx 0.43$, which are close to the obtained values. The coefficients $\at$ and $\atti$ do not depend on the electron temperature gradient and are uniquely determined for any given collision frequency. We find that the numerically obtained eigen-frequencies are consistent with the dispersion relation for the collisional microtearing mode \eqref{eq:mt_disp} using these coefficients.

The behavior of microtearing instability is explained in terms of the time-dependent thermal/inertia force. In the collisional limit $\nu_{\ele\ion}\rightarrow \infty$, the coefficient $\atti$ becomes constant. Because the time-dependent thermal/inertia force is proportional to $\nu_{\ele\ion}^{-1}$ (see \eqref{eq:mt_mode_freq}), only the thermal force remains. In this limit, the simulation results demonstrate that the microtearing is stable (see Fig.~\ref{fig:nuscan}). In the collisionless limit $\nu_{\ele\ion}\rightarrow 0$, it is clear that the time-dependent thermal/inertia force vanishes as the force is generated by electron--ion collisions. The microtearing mode is also stable in this limit, as shown in Fig.~\ref{fig:nuscan}.
In the $\nu_{\ele\ion}\sim \omega_{\ast\ele}$ regime, the coefficient $\atti$ depends on the collision frequency but is finite; therefore, it can lead to an unstable microtearing mode.
It is uncertain at what collision regime the coefficient $\atti$ vanishes, and the same is true for the existence of a collisional microtearing mode.
This existence is subtle because the coefficient varies depending on the collision model. (Remember that energy diffusion, which is not considered in some theoretical models, strongly suppresses the growth of the mode.) 
The present result quantitatively demonstrates the existence of the microtearing mode unstable regime using a comprehensive collision model.

\subsection{Energy evolution on the microtearing mode}
We perform a nonlinear simulation to show the energy evolution of the microtearing mode.
The generalized energy $W$ consisting of the particle part $K_s$ and magnetic field part $M_{\perp,\para}$ for the gyrokinetic equation is given as
\begin{align}
	W &= \sum_{s} K_{s} + M_{\perp} + M_{\para} \\ \nonumber
	&= \int \left[ 
	\sum_{s} \left(
	    \frac{n_{0s} T_{0s}}{2}\frac{\tilde{n}_{s}^2}{n_{0s}^2}
		+ \frac{m_s n_{0s} \tilde{u}_{\para,s}^2}{2} 
		+ \int \frac{T_{0s} \tilde{h}_{s}^{'}}{2 f_{0s}} d\bm{v}
		\right)
	+ \frac{|\nabla_{\perp} \tilde{A}_{\para}|^2}{2\mu_{0}} + \frac{|\delta \tilde{B}_{\para}|^{2}}{2\mu_{0}}
	\right] \diff \bm{r},
\end{align}
where $\tilde{n}_{s}$ and $\tilde{u}_{\parallel,s}$ are the perturbations of the density and parallel flow of the species $s$ calculated from the perturbed distribution function $\delta \tilde{f}_{s}$. The generalized energy evolves as $\diff W/\diff t = \sum_{s} (I_{s} - D_{s})$, where $I_{s}$ is the energy injection associated with the background gradients, and $D_{s}\geq 0$ is the collisional dissipation. The perturbed distribution function is decomposed as in~\cite{zocco2011},
\begin{align}
    \delta \tilde{f}_{s} = \left( \frac{\tilde{n}_{s}}{n_{0s}} + \frac{2v_{\para} \tilde{u}_{\para,s}}{v_{\mathrm{th},s}^2}\right) f_{0s} + \tilde{h}^{'}_{s}.
\end{align}
We also consider the perpendicular flow velocity
\begin{equation}
    \tilde{\bm{u}}_{\perp,s} = \frac{1}{n_{0s}}\int \bm{v}_{\perp} \delta \tilde{f}_{s} \diff \bm{v}
\end{equation}
and associated kinetic energy $\int m_{s}n_{0s} |\tilde{\bm{u}}_{\perp,s}|^2/2 \diff \bm{r}$. 

\begin{figure}[htbp]
\begin{center}
	\includegraphics[width=120mm]{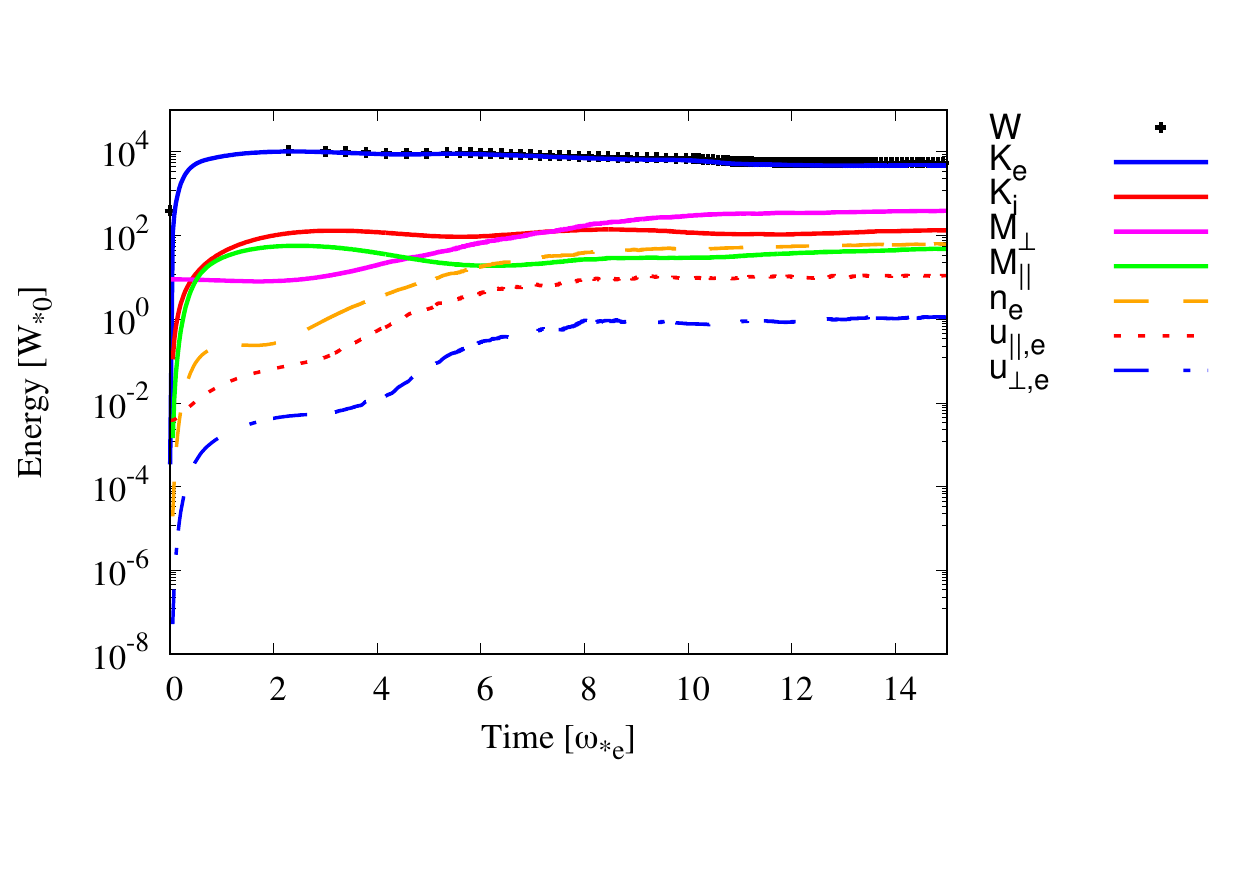}
\end{center}
	\caption{Energy evolution of the microtearing mode. All parameters are chosen to be the same as those in the most unstable case, with the energy diffusion shown in Fig.~\ref{fig:nuscan}. Although the energy of electrons $K_{\ele}$ almost dominates the total energy $W$, its component is not supported by either the density perturbation $n_{\ele}$ or the parallel $u_{\para,\ele}$ and perpendicular $u_{\perp,\ele}$ flow velocities. The magnetic energy $M_{\perp}$ remains at very low level and has little influence on the microtearing mode.}
\label{fig:energy}
\clearpage
\end{figure}

Figure~\ref{fig:energy} shows the time evolution of energy for the microtearing mode. 
The black symbols correspond to the total energy, the blue and red lines are the energies of the electrons and ions, respectively, and the magenta and green lines are the perpendicular and parallel magnetic-field energies, respectively. The yellow, red, and blue dashed lines correspond to the energy of the electrons associated with the density perturbation and parallel and perpendicular flows, respectively.
Each energy is normalized by the energy defined by the diamagnetic drift velocity
\begin{equation}
    W_{0\ast} \equiv \int \frac{1}{2}m_{\ele}n_{0\ele} v_{0\ast}^{2} \diff \bm{r}
    = V \frac{n_{0\ele}T_{0\ele}}{4}
        \left( \frac{\rho_{\ele}}{L_{T_{0\ele}}} + \frac{\rho_{\ele}}{L_{n_{0\ele}}} \right)^{2},
\end{equation}
where $v_{0\ast} = (\omegat+\omegan)/k_{y}$, and $V$ is the total volume of the domain. 

Because the system is driven by the temperature gradient, the total energy initially increases, eventually saturating where the energy injection $\sum_{s} I_{s}$ and dissipation $\sum_{s} D_{s}$ balance.
We found via preliminary analyses that the electron energy, which accounts for the majority of the energy, is not the energy associated with either the density perturbation or the parallel and perpendicular flow velocities and may be supported by the energy owing to the electron temperature perturbation. The further detailed analyses will be investigated in the future.
During the process, the magnetic energy remains at a very low level and has little influence on the process.
This result is in marked contrast to the magnetic reconnection process driven by the tearing mode presented in~\cite{numata2015}, where the initial magnetic energy is drastically released to heat plasmas.
Unlike in the normal tearing mode, an explosive release of magnetic energy does not occur in the microtearing mode. We conclude that the microtearing and normal tearing modes are completely different processes, as indicated by their driving mechanisms.

\section{Summary}
\label{sec:summary}
In this papar, we have presented theoretical and numerical studies on the destabilization mechanism of the collisional microtearing mode.

We have reviewed linear theoretical analyses in a slab geometry based on Hassam’s fluid equations and have shown that the essential destabilizing mechanism of the mode is the phase lag of the inductive electric field behind the perturbed magnetic field. The lag is induced by the time-dependent thermal force and/or collisional inertia force. 
Because it is not magnetically driven, a similar instability exists even without magnetic shear (the so-called self-filamentation mode).

We have perfomed linear gyrokinetic simulations using $\agk$ code to investigate the fundamental properties of the microtearing mode. The microtearing mode becomes unstable in the regime where $\nu_{\ele\ion} \sim \omega_{\ast \ele}$ and is stable in collisionless and collisional limits. By quantitatively estimating the coefficients of the time-dependent thermal/inertia force, we have obtained the dispersion relation of the microtearing mode, which agrees well with the theoretical prediction. The unstable regime is identified as $0.01 < \nu_{\ele\ion}/\omega_{\ast \ele} < 1$.

We infer the implications of our results to the present and future experiments.
We refer to the experimental data on the NSTX device. 
According to Table 1 in~\cite{guttenfelder2012}, at \RN{$r/r_{0}=0.5$} ($r$ and $r_{0}$ are the radial coordinate and minor radius of the device, respectively), 
\RN{the collision frequency normalized by the diamagnetic frequency $\nu_{\mathrm{ei}}^{\mathrm{ex}}/\omega_{\ast\mathrm{e}}$ is estimated as $\nu_{\mathrm{ei}}^{\mathrm{ex}}/\omega_{\ast\mathrm{e}}\ \approx 0.878$.} 
The poloidal wavenumber $k_\theta \rho_\mathrm{s}=0.6$ is used, which can dominantly drive microtearing instability. 
\RN{Using the electron temperature of ITER shown in \cite{sips2005} to estimate the collision frequency, $\nu_{\ele\ion}/\omega_{\ast\ele} \approx 0.0281$ because of $\nu_{\ele\ion} \propto T_{\ele}^{-3/2}$.}
Our result in Fig.~\ref{fig:nuscan} indicates that \RN{ITER and NSTX} experiments are in an unstable regime. 
It has been also shown that the collisional microtearing mode is independent of $\beta$. We expect the same level of instability in future higher-$\beta$ devices. 
\RN{Note that $\beta$ dependence of collisionless microtearing modes driven by other mechanisms, such as trapped particles are debatable.}

We have also presented a nonlinear simulation result showing the energy evolution for the microtearing mode. 
Although the energy associated with the electron temperature perturbation may be dominant in the microtearing mode, the detailed analyses are left behind as the future works.
No explosive magnetic energy release is observed.

The collisionless microtearing mode is not observed in the present study. 
\MY{
We have derived a condition for stabilization of the microtearing mode in the collisionless regime although it is not a proof of non-existence of the collisionless microtearing mode (see \eqref{eq:stabilization_condition_collisionless_mt}).
Careful investigation of the existence of unstable modes in the purely collisionless regime remains as future work.
}

As discussed in this paper, phase lag is essential for the destabilization of the microtearing mode, similar to the destabilization of electrostatic drift waves. The instability can be discussed in terms of the mechanism leading to the phase lag. 
In toroidal geometry, trapped particles may provide a phase lag and lead to instability in the collisionless regime.
\RN{The stability of the collisionless electromagnetic tearing-parity ETG mode is also \MY{an} important issue although it is not discussed in this work. As shown in~\cite{geng2020}, the electron finite Larmor radius (FLR) effect is a key component, which is not considered in our present simulation. The relationship between the destabilization mechanism of the microtearing mode and the collisionless electromagnetic tearing-parity ETG mode may be discussed elsewhere.}

\section*{Acknowledgements}
This work was supported by JSPS KAKENHI Grant Number 22K03568.
The computation in this study was performed at the facilities of the Center for Cooperative Work on Computational Science, University of Hyogo, the JFRS-1 supercomputer system at the Computational Simulation Centre of the International Fusion Energy Research Centre (IFERC-CSC) at the  Rokkasho Fusion Institute of QST (Aomori, Japan) and on the 'Plasma Simulator' (NEC SX-Aurora TSUBASA) of NIFS with the support and under the auspices of the NIFS Collaboration Research program (NIFS22KISS019).
We would like to thank Editage [\url{http://www.editage.com}] for editing and reviewing this manuscript for English language.


\section*{References}
\bibliography{bib/yagyu2022}

\bibliographystyle{iopart}  
\end{document}